\documentclass[12pt]{article}
\pdfoutput=1
\usepackage{epsfig}
\usepackage{amsfonts}
\usepackage{amssymb}
\usepackage{xcolor}
\usepackage{titlesec}
\usepackage{multirow}
\usepackage{bm}
\usepackage{amsmath}
\usepackage{comment}
\usepackage{silence}
\titleformat{\paragraph}
{\normalfont\normalsize\bfseries}{\theparagraph}{1em}{}
\titlespacing*{\paragraph}
{0pt}{3.25ex plus 1ex minus .2ex}{1.5ex plus .2ex}
\begin{document}
\newcommand{\nc}{\newcommand}
\nc{\beq}{\begin{equation}} \nc{\eeq}{\end{equation}}
\nc{\beqa}{\begin{eqnarray}} \nc{\eeqa}{\end{eqnarray}}
\nc{\eps}{{\epsilon}}
\nc{\R}{{\cal R}}
\nc{\A}{{\cal A}}
\nc{\K}{{\cal K}}
\nc{\B}{{\cal B}}
\nc{\C}{{\cal C}}
\nc{\N}{{\cal N}}
\begin{center}

{\bf \Large   High Energy Behaviour in Maximally\\[0.4cm] Supersymmetric Gauge Theories in Various \\[0.4cm] Dimensions } \vspace{1.0cm}

{\bf \large D. I. Kazakov$^{1,4}$, L. V. Bork$^{2,3}$, A. T. Borlakov$^{1,4}$, \\[0.4cm] D. M. Tolkachev$^{1,5}$, D. E. Vlasenko$^{6}$ \\[0.3cm] }\vspace{0.5cm}

{\it
$^1$Bogoliubov Laboratory of Theoretical Physics, Joint
Institute for Nuclear Research, Dubna, Russia.\\
$^2$Alikhanov Institute for Theoretical and Experimental Physics, Moscow, Russia\\
$^3$Center for Fundamental and Applied Research, All-Russian Institute of Automatics, Moscow, Russia \\
$^4$Moscow Institute of Physics and Technology, Dolgoprudny, Russia\\
$^5$Stepanov Institute of Physics, Minsk, Belarus\\
$^6$Department of Physics, South Federal State University, Rostov-Don, Russia}
\vspace{0.5cm}

\abstract{Maximally supersymmetric field theories in various dimensions are believed to possess special properties due to  extended supersymmetry. In four dimensions they are free from UV divergences but are IR divergent on shell, in higher dimensions, on the contrary, they are IR finite but UV divergent. In what follows we consider the four-point on-shell scattering amplitudes in D=6,8,10 supersymmetric Yang-Mills theory in the planar limit within  the spinor-helicity  and on shell supersymmetric formalism. We study the UV divergences  and demonstrate how one can sum them over all orders of PT. Analyzing the $\R$-operation, we obtain the  recursive relations and derive differential equations that sum  all leading, subleading, etc., divergences in all loops  generalizing the standard RG formalism for the case of nonrenormalizable interactions. We then perform the renormalization procedure which differs from the ordinary one in that the renormalization constant becomes the operator depending on kinematics. Solving the obtained  RG equations for particular sets of diagrams analytically and for the general case numerically, we analyze their high energy behaviour and find out that while each term of PT increases as a power of energy, the total sum behaves differently:  in D=6 two partial amplitudes decrease with energy and the third one increases  exponentially, while in D=8 and 10 the amplitudes possess an infinite number of periodic poles at finite energy.}
\end{center}


\section{Introduction}

In the last decade we witnessed serious progress in understanding the structure of  the
amplitudes (the S-matrix) in gauge theories in various dimensions (for review see, for example,
\cite{Reviews_methods,BDS4point3loop_et_all,N=8SUGRA finiteness}). This progress became
possible due to the development of new  techniques: the spinor helicity and momentum twistor
formalisms
\cite{Reviews_methods,GeneralDimensions}, different sets of recurrence relations for the tree level amplitudes, the unitarity 
based methods for the loop amplitudes \cite{Reviews_methods} and various realizations of the
on-shell superspace formalism for theories with supersymmetry \cite{Reviews_Ampl_General}. 

The subject of investigation was mainly related to the so-called maximally supersym\-metric
theories, which are believed to possess special properties due to higher symmetries. The
number of supersymmetries $\N$ that can be realized in D dimensions is limited if one restricts
the maximal spin of the states. For the gauge theories with maximal spin 1 one has  the
following  maximally supersym\-metric theories:
$ D=4 \ \mathcal{N}=4,  D=6 \ \mathcal{N}=2,  D=8 \ \mathcal{N}=1,  D=10 \ \mathcal{N}=1.$

While the $\mathcal{N}=4$ SYM theory is completely  on shell  UV finite and possesses  only the
IR divergences, in higher dimensions the situation is the opposite: there are no IR divergences
even on shell but all theories are UV nonrenormalizable by power counting. 
Among D=4 gauge theories  $\mathcal{N}=4$ SYM \cite{Reviews_Ampl_General} possesses some
exceptional properties and is expected to be exactly solvable. So one could expect that 
$\mathcal{N}=4$ higher dimensional counterparts will also be, in some sense, exceptional
theories. 

Indeed, the integrands of the four-point amplitudes in SYM theory in any even dimension have
almost identical form (only the tree level amplitudes, which are the common factors, are
different). This is the consequence of the dual (super)conformal symmetry which is present, in some form, in all the above mentioned SYM theories \cite{GeneralDimensions,SpinorHelisity_extraDimentions}. 

In the sequence of papers~\cite{we0,we1,we2,we3}, we considered the leading and subleading UV
divergences of the on-shell four point scattering amplitudes for all three cases of maximally
supersymmetric SYM theories, D=6 (N=2 SUSY), D=8 (N=1 SUSY) and D=10 (N=1 SUSY). We obtained
the recursive relations that allow one to get leading and subleading divergences in all
loops in a pure algebraic way \cite{we2,we3}. Then we constructed the differential equations, which are the
generalization of the RG equations for non-renormalizable theories \cite{we2,we3}. Similar to the
renormalizable theories, these equations lead to summation of the leading (and subleading)
divergences in all loops. In \cite{we4}, we concentrated on solving these equations.  For a
particular set of diagrams these equations allow for an analytical solution while in the
general case we applied numerical methods. Remarkably, numerical solutions follow the general
pattern of their analytical
counterparts, which allows one to trace the main properties of the solutions explicitly.  In
\cite{we4}, we considered  also the sub-subleading case and focused on the scheme dependence of
the counter terms. We studied the transition from the minimal to non-minimal subtraction scheme
and showed that it was equivalent to the redefinition of the dimensionless couplings $g^2s^2$
or $g^2t^2$ similar to the renormalizable case. The difference from the latter case manifests
itself in the fact that the renormalization constant becomes the operator depending on the
kinematics. The peculiarities of the renormalization procedure for higher dimensional theories
is discussed in detail in \cite{we5}. 

In this paper,  we summarize all our results in studying maximally supersymmetric gauge
theories in $D=6,8,10$ dimensions. For the sake of completeness, in Sec.2, we remind the spinor
helicity and on shell superspace formalism. In Sec.3, we consider the diagrams for four-point 
color-ordered planar scattering amplitudes which appear in this formalism and analyze their UV
properties. We  explain how the $\R$-operation works and derive the recursive relations which
allow one to get the all-loop expressions for the leading, subleading, etc., divergences.  We
then convert these relations into differential equations which are the generalization of the
familiar RG equations for the case of non-renormalizable interactions. In Sec.4, we consider
the properties of these equations and solve them
analytically for particular sets of diagrams and numerically in the general case. Section 5 is
dedicated to the renormalization procedure. We show that it reminds the usual one when the UV
divergences are removed by multiplication by the renormalization constants resulting in
multiplicative renormalization of the coupling, but in this case it is not a simple
multiplication but rather the action of the renormalization operator. Finally, in Sec.6, we
discuss the all-loop high energy asymptotics of the four-point scattering amplitudes in
$D=6,8,10$ dimensions. The last section contains the summary of our views and conclusions.

\section{
Spinor-helicity formalism in various dimensions and amplitudes in D=6,8,10 SYM theories}

\subsection{Spinor-helicity formalism}
As was mentioned in the introduction, the spinor helicity
and the on shell momentum superspace formalisms play a
crucial role in the above mentioned  achievements in understanding the structure of the S-matrix of four dimensional supersymmetric gauge field theories. In the following two sections we will discuss essential details of both formalisms. Let us start with the generalization of the spinor helicity formalism to the case of even
dimensions $D=6,8$ and $10$. In our discussion we manly follow \cite{GeneralDimensions}. The corresponding the on-shell momentum superspace  will be discussed in the next section.

In even dimensions one can always choose the chiral representation of the gamma matrices as $\Gamma^{\mu}$ (as usual $\{\Gamma^{\mu},\Gamma^{\nu}\}=2\eta^{\mu\nu}$):
\begin{equation}
 \Gamma^{\mu}=\left(
\begin{array}{ccc}
  0 & (\sigma^{\mu})^{AB'} \\
 (\overline{\sigma}^{\mu})_{B'A} & 0 \\
 \end{array}
\right).
\end{equation}
Here $\mu$ is the $SO(D-1,1)$ Lorentz group in $D$ dimensions vector representation index, $A$ and $B'=1,...,2^{D/2-1}$ are the $Spin(SO(D-1,1))$ indices. Here we will be interested in $D=4,6,8,10$. The explicit form of $(\sigma^{\mu})^{AB'}$ and $(\overline{\sigma}^{\mu})_{B'A}$ can be found in \cite{GeneralDimensions}. Using this notation, one can decompose the Dirac spinor $\psi$ as a pair of Weyl chiral and anti-chiral spinors $\lambda^A$ and $\tilde{\lambda}_{A'}$.

One can construct the Lorentz invariant $\psi_1^TC\psi_2$ from two Dirac spinors
$\psi_2$ and $\psi_1$ using the charge conjugation matrix $C$ defined so that
\begin{eqnarray}
C\Gamma^{\mu}C^{-1}=-(\Gamma^{\mu})^T.
\end{eqnarray}
The explicit form of $C$ can be found in \cite{GeneralDimensions}.
As for the Weyl spinors, there are two possible decompositions of $C$ depending on dimension:
\begin{equation}
 C=\left(
\begin{array}{ccc}
  \Omega_{BA} & 0 \\
 0 & \Omega^{B'A'} \\
 \end{array}
\right),
\end{equation}
and
\begin{equation}\label{OmegaD610}
 C=\left(
\begin{array}{ccc}
  \Omega_{B}^{A'} & 0 \\
 0 & \Omega^{B'}_{A} \\
 \end{array}
\right)
\end{equation}
respectively, for $D=4,8$ and $D=6,10$. The $\Omega$ matrices obey the following relations:
\begin{equation}
  \Omega_{BA} \Omega^{AC}=\delta^C_B,~ \Omega_{B'A'} \Omega^{A'C'}=\delta^{B'}_{C'},
\end{equation}
for $D=4,8,$ and
\begin{equation}
\Omega_B^{A'}\Omega_{A'}^C=\delta^C_B,~\Omega_A^{B'}\Omega_{C'}^A=\delta^{B'}_{C'}
\end{equation}
for $D=6,10$.

One can use the matrices $\Omega$ to raise and lower the indices of the spinors
\begin{eqnarray}
\lambda_A=\lambda^B\Omega_{BA}~\tilde{\lambda}^{A'}=\Omega^{B'A'}\tilde{\lambda}_{B'}
\end{eqnarray}
for $D=4,8$ and to relate the chiral and antichiral spinors
\begin{eqnarray}
\lambda_A=\Omega^{A'}_A\tilde{\lambda}_{A'},~\tilde{\lambda}^{A'}=\lambda^A\Omega_A^{A'}
\end{eqnarray}
for $D=6,10$. 
Using these properties, one can also construct the Lorentz invariants for the pair of spinors which are labeled  by $i$ and $j$ in two ways:
\begin{eqnarray}
\lambda^B_i\Omega_{BA}\lambda^A_j\equiv \langle ij \rangle,
~\tilde{\lambda}_{B',i}\Omega^{B'A'}\tilde{\lambda}_{A',j}\equiv[ ij ]
\end{eqnarray}
and
\begin{eqnarray}
\tilde{\lambda}_{A',i}\Omega^{A'}_A\lambda^A_j\equiv[i|j\rangle,
~\lambda^A_i\Omega_A^{A'}\lambda_{A',j}\equiv\langle i|j]
\end{eqnarray}
for $D=4,8$ and $D=6,10$, respectively. 
The matrices $C$ can be always chosen in such a way that
\begin{eqnarray}
C^T&=&-C~\mbox{for}~D=4,10,\nonumber\\
C^T&=&C~\mbox{for}~D=6,8.
\end{eqnarray}

In some dimensions it is also possible to construct additional Lorentz invariants. For example, in $D=6$ one has
$Spin(SO(5,1))\cong SU(4)^*$, so one can use absolutely antisymmetric tensor $\varepsilon_{ABCE}$ asso\-siated with $SU(4)^*$ to contract spinorial indices of four spinors:
$\epsilon_{ABCD}\lambda^{A}_1\lambda^{B}_2\lambda^{C}_3\lambda^{D}_4
  \equiv \langle 1234 \rangle$, $\epsilon^{ABCD}\tilde{\lambda}_{A,1}\tilde{\lambda}_{B,2}\tilde{\lambda}_{C,3}\tilde{\lambda}_{D,4}\equiv [ 1234 ]$.
These combinations are also Lorentz invariant.

It is always possible to relate light like (massless) momentum $p_{\mu}$ with the pair of Weyl spinors using the  Dirac equations for the spinors $\lambda^A$ and $\tilde{\lambda}_{A'}$:
\begin{eqnarray}
  (p_{\mu}\sigma^{\mu})^{BA'}\tilde{\lambda}_{A'}=0~\mbox{and}~
  (p_{\mu}\tilde{\sigma}^{\mu})\lambda^A=0.
\end{eqnarray}
The solutions to these equations can be labeled by additional helicity indices
$a$ and $a'$ which transform under the little group of the Lorentz group, which is
$SO(D-2)$ in our  case.  We want to stress that in $D > 4$ dimensions helicity of a massless particle is no longer conserved and transforms according to the little group similarly to helicity of a massive particle in $D=4$.
From the Dirac equations one can see that
\begin{eqnarray}
  (p_{\mu}\sigma^{\mu})^{BA'}\tilde{\lambda}_{A'a'}=0,~
  (p_{\mu}\tilde{\sigma}^{\mu})\lambda^{Aa}=0,
\end{eqnarray}
and for their conjugates
\begin{eqnarray}
  (p_{\mu}\sigma^{\mu})^{BA'}\lambda_{B}^{a'}=0,~
  (p_{\mu}\tilde{\sigma}^{\mu})\tilde{\lambda}^{A'}_a=0.
\end{eqnarray}
One can take the solutions to these equations
$\tilde{\lambda}_{A'a'}(p),\lambda^{Aa}(p)$ (and their conjugates) in
such a way that
\begin{eqnarray}\label{SpinorToMomentum}
  \sum_{a}\lambda^{Ba}(p)\tilde{\lambda}^{A'}_{a}(p)=p_{\mu}(\sigma^{\mu})^{BA'},~
  \sum_{a'}\tilde{\lambda}_{B'a'}(p)\lambda^{a'}_{A}(p)=p_{\mu}(\tilde{\sigma}^{\mu})_{B'A}.
\end{eqnarray}
This gives us the desired representation of light like momentum $p_{\mu}$ as a pair of Weyl spinors.

Using Weyl spinors, one can also construct a representation for the polarization vectors of gluons in $D$ dimensions, which is given up to normalisation by 
\begin{eqnarray}
  \varepsilon^{\mu}_{aa'}(p|q)= q_{\nu}\frac{\tilde{\lambda}_{a}(p)(\overline{\sigma}^{\mu}\sigma^{\nu})\tilde{\lambda}_{a'}(q)}{(pq)},~
  \varepsilon^{\mu,aa'}(p|q)= q_{\nu}\frac{\lambda^{a}(p)(\sigma^{\mu}\sigma^{\nu})\lambda^{a'}(q)}{(pq)}.
\end{eqnarray}
Note that the polarization vectors contain the dependence on an additional parameter (vector) $q$. This dependence parametrizes the  gauge ambiguity and the dependence on $q$ must cancel in gauge invariant objects such as scattering amplitudes. 
The polarization vectors for massless fermions can be chosen as Weyl spinors while the polarization vectors for scalars are trivial.

Using this  representation of momenta and polarisation vectors in terms of Weyl spinors one can always write down the scattering amplitude in the gauge theory in arbitrary even dimension, which is the function of the Lorentz invariant products of momenta and polarization vectors in terms of the spinor products corresponding to the momenta of external particles only.

\subsection{On-shell momentum superspace}
In this section, we will discuss the essential details regarding the on shell momentum superspace
constructions in  $D=6,8,10$ dimensions.

With the on shell momentum superspace one can obtain a compact representation for
the amplitudes (all amplitudes with different particles are combined in a single object) in supersymmetric gauge theories,
which is very  convenient in the unitarity based computations \cite{Reviews_methods}.

Let us start with the $D=6$ $\mathcal{N}=(1,1)$
on-shell momentum superspace formalism first \cite{Sigel_D=6Formalism}.
The on-shell
$\mathcal{N}=(1,1)$  superspace for $D=6$  can be parameterized by the following set of coordinates:
\begin{eqnarray}\label{Full_(1,1)_superspace}
  \mbox{$\mathcal{N}=(1,1)$ D=6 on-shell superspace}=\{\lambda^A_a,\tilde{\lambda}_{A}^{\dot{a}},\eta_a^I,\overline{\eta}_{I'\dot{a}}\},
\end{eqnarray}
where $\eta_a^I$ and $\overline{\eta}_{\dot{a}}^{I'}$ are the Grassmannian coordinates,
$I=1,2$ and $I'=1',2'$ are the $SU(2)_R\times SU(2)_R$ R-symmetry indices. Note that
this superspace is not chiral. In this superspace one has two types of supercharges $q^{A I}$ and
$\overline{q}_{A I'}$ with the commutation relations
\begin{eqnarray}\label{commutators_for_superchrges_full_(1,1)}
  \{ q^{A I}, q^{B J}\}&=&p^{AB}\epsilon^{IJ},\nonumber\\
  \{ \overline{q}_{A I'}, \overline{q}_{B J'}\}&=&p_{AB}\epsilon_{I'J'},\nonumber\\
  \{ q^{A I},  \overline{q}_{B J'}\}&=&0.
\end{eqnarray}

Using this superspace, similar to the $D=4$ SYM case \cite{Reviews_methods}, one can combine all creation/annihilation operators of the on shell states from the $\mathcal{N}=(1,1)$ supermultiplet into a single combination which is invariant under on-shell SUSY transformations. The $\mathcal{N}=(1,1)$ supermultiplet itself is given
by the following creation/annihilation operators
$$
\{A_{a\dot{a}},~\Psi^a_I,~\overline{\Psi}^{I'\dot{a}},~\phi^{I'}_I\},
$$
which correspond to the physical polarizations of the gluon $|A_{a\dot{a}}\rangle$, two fermions
$|\Psi^a_I\rangle$,$|\overline{\Psi}^{I'\dot{a}}\rangle$ and two complex scalars
$|\phi^{I'}_I\rangle$ (antisymmetric with respect to $I,I'$). This multiplet is CPT
self-conjugated. 

However, to do this, one has to perform a truncation of the full
$\mathcal{N}=(1,1)$ on-shell superspace \cite{Sigel_D=6Formalism}. This can be done consistently
by using a special version of harmonic superspace \cite{Sigel_D=6Formalism}.
The harmonic variables $u_{I}^{\mp}$ and $\overline{u}^{\pm
I'}$ in this setup must be chosen to parameterize the double coset space
\begin{eqnarray}
\frac{SU(2)_R}{U(1)}\times\frac{SU(2)_R}{U(1)}.
\end{eqnarray}
Using these variables, we express the projected supercharges and the Grassmannian coordinates as
\begin{eqnarray}
  q^{\mp A}&=&u^{\mp}_Iq^{A I},~
  \overline{q}^{\pm}_{A}=u^{\pm I'}\overline{q}_{A I'},
  \nonumber\\
  \eta^{\mp}_a&=&u^{\mp}_I\eta_a^I,~
  \overline{\eta}^{\pm}_{\dot{a}}=u^{\pm I'}\overline{\eta}_{I'\dot{a}}.
\end{eqnarray}
We can also reexpress all creation/annihilation operators of the on-shell states of $D=6$ $\mathcal{N}=(1,1)$ SYM using our new harmonic variables. The bosonic states are
\begin{equation}\label{(1,1)_onshel_states bos}
\phi^{--},~\phi^{-+},~\phi^{+-},~\phi^{++},A^{a\dot{a}},
\end{equation}
while the fermionic states are
\begin{equation}\label{(1,1)_onshel_states ferm}
\Psi^{-a},~\Psi^{+a},~\overline{\Psi}^{-\dot{a}}~\overline{\Psi}^{+\dot{a}}.
\end{equation}
Then we have to consider only the objects that depend on the set of
variables that parameterize the subspace ("analytic superspace") of
the full $\mathcal{N}=(1,1)$ on-shell superspace
\begin{eqnarray}\label{Truncated_onshell_superspace}
  \mbox{$\mathcal{N}=(1,1)$ D=6 on-shell harmonic  superspace}=\{\lambda^A_a,\tilde{\lambda}_{A}^{\dot{a}},
\eta^{-}_a,\overline{\eta}_{\dot{a}}^{+} \}.
\end{eqnarray}
The projected supercharges and momentum generators acting on the analytic superspace
for the n-particle case can be explicitly written as:
\begin{eqnarray}\label{projected_supercharges_n_particle_state}
  p^{AB}=\sum_i^n\lambda^{Aa}(p_i)\lambda^B_{a}(p_i),~ q^{-A}=\sum_i^n\lambda^A_a(p_i)\eta^{-a}_i,~
  \overline{q}_A^+=\sum_i^n\tilde{\lambda}_A^{\dot{a}}(p_i)\overline{\eta}_{\dot{a},i}^+.
\end{eqnarray}
Now one can finally combine all the on-shell state creation/annihilation operators
(\ref{(1,1)_onshel_states bos} and \ref{(1,1)_onshel_states ferm}) into
one superstate $|\Omega_i\rangle=\Omega_i|0\rangle$ (here $i$ labels the momenta
carried by the state):
\begin{eqnarray}
  |\Omega_i\rangle&=&\{ \phi^{-+}_i+\phi^{++}_i(\eta^-\eta^-)_i+
  \phi^{--}_i(\overline{\eta}^+\overline{\eta}^+)_i
  +\phi^{+-}_i(\eta^-\eta^-)_i(\overline{\eta}^+\overline{\eta}^+)_i
  \nonumber\\
  &+&(\Psi^+\eta^-)_i+(\overline{\Psi}^-\overline{\eta}^+)_i+
  (\Psi^-\eta^-)_i(\overline{\eta}^+\overline{\eta}^+)_i+
  (\overline{\Psi}^+\overline{\eta}^+)_i(\eta^-\eta^-)_i\nonumber\\
  &+&(A\eta^-\overline{\eta}^+)_i\}|0\rangle,
\end{eqnarray}
where $(XY)_i\doteq X^{a/\dot{a}}_iY_{i~a/\dot{a}}$. Hereafter we
will drop the $\pm$ labels for simplicity.
As in the $D=4$ case, we can formally write the colour ordered amplitude as
\begin{eqnarray}
 \mathcal{A}_n(\{\lambda^A_a,\tilde{\lambda}_{A}^{\dot{a}},
\eta_a,\overline{\eta}_{\dot{a}} \})=\langle
0|\prod_{i=1}^n\Omega_i S|0\rangle,
\end{eqnarray}
where $S$ is the S-matrix operator of the theory, the average $\langle0| \ldots |0\rangle$ is understood with
respect to some (for example, component) formulation of the theory.
The invariance with respect to translations and supersymmetry transformations
requires the amplitude to be annihilated by the corresponding generators
\begin{eqnarray}
 p^{AB}\mathcal{A}_n=q^A\mathcal{A}_n=\overline{q}_A\mathcal{A}_n=0.
\end{eqnarray}
Thus, the superamplitude should have the form
\begin{eqnarray}
 \mathcal{A}_n(\{\lambda^A_a,\tilde{\lambda}_{A}^{\dot{a}},
\eta_a,\overline{\eta}_{\dot{a}} \})=
\delta^6(p^{AB})\delta^4(q^A)\delta^4(\overline{q}_A)\mathcal{P}_n(\{\lambda^A_a,\tilde{\lambda}_{A}^{\dot{a}},
\eta_a,\overline{\eta}_{\dot{a}} \}),
\end{eqnarray}
where $\mathcal{P}_n$ is a polynomial with respect to
$\eta$ and $\overline{\eta}$ of degree  $2n-8$. Since  helicity  is not conserved any more (is not a conserved quantum number) in contrast to the $D=4$ case, there are no closed subsectors of MHV, NMHV, etc. amplitudes.

The Grassmannian  delta functions $\delta^4(q^A)$ and
$\delta^4(\overline{q}_A)$ in the case under consideration are defined as
\begin{eqnarray}
  \delta^4(q^A)&=&\frac{1}{4!}\epsilon_{ABCD}
  \hat{\delta}(q^A)\hat{\delta}(q^B)
  \hat{\delta}(q^C)\hat{\delta}(q^D),\nonumber\\
  \delta^4(\overline{q}_A)&=&\frac{1}{4!}\epsilon^{ABCD}
  \hat{\delta}(\overline{q}_{A})\hat{\delta}(\overline{q}_{B})
  \hat{\delta}(\overline{q}_{C})\hat{\delta}(\overline{q}_{D}).
\end{eqnarray}
The delta function $\hat{\delta}(X^I)$ here is the usual  Grassmannian delta function defined as $\hat{\delta}^N(X^I) \equiv \prod_{I=1}^N X^I$, where $I$ is the R-symmetry index. In the harmonic formulation we simply have $\hat{\delta}(X) \equiv X$.

Let us consider now the four-point amplitude.
The degree of  the Grassmannian polynomial $\mathcal{P}_4$ is
$2n-8=0$, so $\mathcal{P}_4$ is a function of bosonic variables
$\{\lambda^A_a,\tilde{\lambda}_{A}^{\dot{a}}\}$ only, just as in the $D=4$ case
\begin{eqnarray}\label{4-point_ampl_general_form}
  \mathcal{A}_4(\{\lambda^A_a,\tilde{\lambda}_{A}^{\dot{a}},
\eta_a,\overline{\eta}_{\dot{a}} \})=
  \delta^6(p^{AB})\delta^4(q^A)\delta^4(\overline{q}_A)
  \mathcal{P}_4(\{\lambda^A_a,\tilde{\lambda}_{A}^{\dot{a}}\}).
\end{eqnarray}
At the tree level $\mathcal{P}_4$ can be found explicitly from a comparison with the expression for the 4-gluon amplitude \cite{Sigel_D=6Formalism}
obtained with the help of the six dimensional version of the BCFW recurrence relation or direct Feynman diagram computation
\cite{SpinorHelisity_extraDimentions}.
This gives us that in fact $\mathcal{P}_4^{(0)}$ has a very simple form: $\mathcal{P}_4^{(0)}\sim1/st$.
Here $s$ and $t$ are the standard Mandelstam
variables defined as $s=(p_1+p_2)^2$ and $t=(p_2+p_3)^2$. We also drop overall coupling constant dependence. 
So one can see that at the tree level the 4-point
superamplitude can be written as:
\begin{eqnarray}\label{4_point_tree_superamplitude}
  \mathcal{A}_4^{(0)}=\delta^6(p^{AB})\delta^4(q^A)\delta^4(\overline{q}_A)\frac{1}{st}.
\end{eqnarray}
Here $p^{AB}$ and $q^A,\overline{q}_A$ are given by (\ref{projected_supercharges_n_particle_state}).
Note that already at the tree level the 5-point amplitude is not so simple
\cite{SpinorHelisity_extraDimentions} compared to the four point case. However, the iterated two particle cuts, which utilize only the tree level four point amplitude, are enough to reconstruct the loop integrands up to three loops. The form of the integrand coincides with the $D=4$ case up to the tree level amplitude (see Fig.\ref{expan}). One can argue \cite{Reviews_Ampl_General} that this property will hold beyond the three-loop level. 

To illustrate how the unitarity cuts work, we consider a simple one loop computation.
\begin{figure}[ht]
 \begin{center}
 %\leavevmode
  \epsfxsize=6cm
 \epsffile{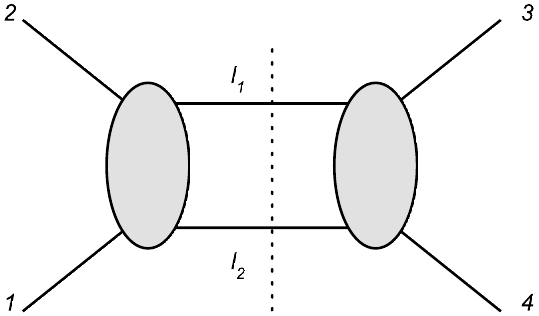}
 \end{center}\vspace{-0.2cm}
 \caption{Two particle s-channel cut for the one loop $D=6$ SYM amplitude.}\label{1loop2particle}
 \end{figure}
The integrand for the  s-channel two particle cut of the one loop amplitude takes the form (here we explicitly label the (super)momentum dependence of the amplitudes) \cite{Sigel_D=6Formalism}
\begin{eqnarray}
  Int\mathcal{A}_4^{(1)}&=&\int d^4\eta_{l_1l_2}d^4\overline{\eta}_{l_2l_1}~
  \mathcal{A}^{(0)}_4(1,2,l_1,l_2)\times \mathcal{A}^{(0)}_4(-l_1,-l_2,3,4)
 \end{eqnarray}
The integrals with respect to the Grassmannian variables
$d^4\eta,d^4\overline{\eta}$ can be evaluated,
and after taking into account momentum conservation conditions, one gets:
(the common factor $g_{YM}^4 N_c$ is omitted)
\begin{eqnarray}\label{2pcutEx}
 Int\mathcal{A}_4^{(1)}
&=&-\int d^4\eta_{l_1l_2}d^4\eta_{l_2l_1}
\frac{\delta^4(q^A_R+q^A_{l_1l_2})\delta^4(q^A_L-q^A_{l_1l_2})
\delta^4(\overline{q}_{A,R}+\overline{q}_{A,l_1l_2})
\delta^4(\overline{q}_{A,L}-\overline{q}_{A,l_1l_2})}{s^2(2+l_1)^2(4+l_2)^2}
\nonumber\\
&=&-\delta^4(q^A_R+q^A_L)\delta^4(\overline{q}_{A,R}+\overline{q}_{A,L})
\frac{2(l_1l_2)^2}{s^2(2+l_1)^2(4+l_2)^2}=\mathcal{A}^{(0)}_4\frac{st}{2}
\frac{-i}{(2+l_1)^2(4+l_2)^2}.\nonumber\\
\end{eqnarray}
The following formula for the Grassmannian integration is useful:
($\int d^2 \eta^a_{l_1}~\int d^2\overline{\eta}^{\dot{b}}_{l_2}
  \equiv \int d^4\eta_{l_1l_2}$)
\begin{eqnarray}\label{Two_particle_Grassmann_int}
  &&\int d^4\eta_{l_1l_2}d^4\eta_{l_2l_1}~
  \delta^4(\lambda^{Aa}_{l_1} \eta_{a,l_1} +\lambda^{Aa}_{l_2} \eta_{a,l_2} + q^A_1)\delta^4(\lambda^{Aa}_{l_1} \eta_{a,l_1} +\lambda^{Aa}_{l_2} \eta_{a,l_2} \eta-q^A_2)
  \nonumber\\&&\times\delta^4(\tilde{\lambda}^{A\dot{a}}_{l_1}\overline{\eta}_{\dot{a},l_1}+\tilde{\lambda}^{A\dot{a}}_{l_2}\overline{\eta}_{\dot{a},l_2}+\overline{q}_{B})
  \delta^4(\tilde{\lambda}^{A\dot{a}}_{l_1}\overline{\eta}_{\dot{a},l_1}+\tilde{\lambda}^{A\dot{a}}_{l_2}\overline{\eta}_{\dot{a},l_2}-\overline{q}_{B})
  \nonumber\\&&=(2!)^44(l_1,l_2)^2\delta^4(q_1^A+q_2^A)\delta^4(\overline{q}_{B,1}+\overline{q}_{B,2}).
\end{eqnarray}
Equation (\ref{2pcutEx}) is consistent with the  following ansatz for
part of the amplitude associated with the s-channel cut
\begin{eqnarray}
  -\mathcal{A}^{(0)}_4~\frac{st}{2}~B(s,t),
\end{eqnarray}
where $B(s,t)$ is the $D=6$ scalar box function.
The t-channel cut gives the same result, so we conclude that the full one loop level
amplitude has the form:
\begin{eqnarray}\label{1-loop 4point ampl scalar int}
    \mathcal{A}_4^{(1)}=-A^{(0)}_4~\frac{g_{YM}^2 N_c}{2}~st~B(s,t),
\end{eqnarray}
where $B(s,t)$ is a one loop box scalar integral (see Fig.\ref{expan}).

Let us now consider the $D=8$ case. It can be constructed along the same lines as the $D=6$ one but in a more straightforward manner.
Here we follow \cite{GeneralDimensions}.
The $\mathcal{N}=1$ $D=8$ on-shell superspace can be parameterized by the following set of coordinates:
\begin{eqnarray}\label{Full_N=1 D=8_superspace}
  \mbox{$\mathcal{N}=1$ D=8 on-shell superspace}=\{\lambda^{Aa},\tilde{\lambda}_{a}^{A'},\eta_a\},
\end{eqnarray}
where $\eta^a$ are the Grassmannian coordinates, $A$ and $A'$ are the $spin(SO(7,1))$ indices and
$a$ is the little group $SO(6)$ index. The R-symmetry
group here is $U(1)_R$  and $\eta^a$ carries the $+1$ charge with respect to $U(1)_R$. This superspace is chiral.

The commutation relations for the supercharges in this case have the form
\begin{eqnarray}\label{commutators_for_superchrges_N=1 D=8}
  \{ q^{A}, \bar{q}^{B'}\}&=&p^{AB}.
\end{eqnarray}
The supercharges in the on-shell momentum superspace representation for the  $n$-particle state are given by
\begin{eqnarray}\label{commutators_for_superchrges_N=1 D=8 details}
   p^{AB'}=\sum_{i=1}^n\lambda^{Aa}(p_i)\tilde{\lambda}_a^{B'}(p_i),
   ~q^{A}=\sum_{i=1}^n\lambda^{Aa}(p_i)\eta_a, ~\bar{q}^{B'}=\sum_{i=1}^n\tilde{\lambda}_a^{B'}(p_i)\frac{\partial}{\partial\eta_a}.
\end{eqnarray}
The creation/annihilation operator for the $\mathcal{N}=1$ $D=8 $ on-shell supermultiplet are
$$
\{A^{a\dot{a}},~\Psi^a,~\overline{\Psi}_{a},~\phi,~\overline{\phi}\},
$$
which corresponds to the  physical polarizations of the gluon $|A^{a\dot{a}}\rangle$,
two fermions $|\Psi^a\rangle$, $|\overline{\Psi}_{a}\rangle$ and two scalars $|\phi\rangle$, $|\overline{\phi}\rangle$. Using on-shell momentum superspace Grassmann coordinates, 
one can straightforwardly combine the creation/annihilation operator into one "superstate" $|\Omega_i\rangle$ similar to the $D=4$
case
\begin{eqnarray}\label{superstate}
|\Omega_{i}\rangle = \left(\phi_i + \eta_a\Psi^a_i +
\frac{1}{2!}\eta_a\eta_b A^{a\dot{a}}_i +
\frac{1}{3!}\eta_a\eta_b\eta_c \varepsilon^{abcd}\overline{\Psi}_{d,i} +
\frac{1}{4!}\eta_a\eta_b\eta_c\eta_d \varepsilon^{abcd}\overline{\phi}_i\right) |0\rangle.
\end{eqnarray}
Here $\varepsilon^{abcd}$ is the absolutely antisymmetric tensor associated with the little group $SO(6)\cong SU(4)$. No additional complications are needed.

Using the arguments identical to the previous discussion, we conclude that the colour ordered superamplitude should have the form:
\begin{eqnarray}
 \mathcal{A}_n(\{\lambda^{Aa},\tilde{\lambda}_{a}^{A'},\eta_a \})=
\delta^8(p^{AB'})\delta^8(q^A)\mathcal{P}_n(\{\lambda^{Aa},\tilde{\lambda}_{a}^{A'},\eta_a \}),
\end{eqnarray}
where $\mathcal{P}_n$ is a polynomial with respect to
$\eta$ and $\overline{\eta}$ of degree $2n-8$,
and the Grassmannian  delta function $\delta^8(q^A)$ is defined
in this case as:
\begin{eqnarray}
  \delta^8(q^A)=\frac{1}{8!}\epsilon_{A_1...A_8}
  \prod_{i=1}^8\hat{\delta}(q^{A_i}),
\end{eqnarray}
Here $\epsilon_{A_1...A_8}$ is the
absolutely antisymmetric tensor associated with the $spin(SO(7,1))$.

For the four-point  amplitude
the degree of the Grassmannian polynomial $\mathcal{P}_4$ is again
$2n-8=0$, so as in the previous cases $\mathcal{P}_4$ is a function of bosonic variables
and one can  again write the four-point amplitude in the form
\begin{eqnarray}\label{4-point_ampl_general_form8}
  \mathcal{A}_4(\{\lambda^{Aa},\tilde{\lambda}_{a}^{A'},\eta_a \})=
  \delta^8(p^{AB'})\delta^4(q^A)
  \mathcal{P}_4(\{\lambda^{Aa},\tilde{\lambda}_{a}^{A'}\}).
\end{eqnarray}
At the tree level $\mathcal{P}_4$ can be found from a comparison
with the explicit result of Feynman diagram computation or from the expression obtained as a field theory limit of the superstring scattering amplitude \cite{GeneralDimensions}. Similar to the previous discussion, one again has $\mathcal{P}_4^{(0)}\sim 1/st$, so that at the tree level the 4-point superamplitude can again be written as:
\begin{eqnarray}\label{4_point D8_tree_superamplitude}
  \mathcal{A}_4^{(0)}=\delta^8(p^{AB})\delta^8(q^A)\frac{1}{st}.
\end{eqnarray}
Using the iterated two particle cuts, this allows one to reconstruct the answer for the four-point amplitude up to three loops. To perform this computation, the following formula for the Grassmannian integration is useful:
\begin{eqnarray}
  &&\int d^4\eta_{l_1}d^4\eta_{l_2}
  \delta^8(\lambda_{l_1}^{Aa}\eta_{a,l_1}+\lambda_{l_2}^{Aa}\eta_{a,l_2}+q_1^A)
  \delta^8(\lambda_{l_1}^{Aa}\eta_{a,l_1}+\lambda_{l_2}^{Aa}\eta_{a,l_2}-q_2^A)\nonumber\\
  &&=(4!)^24(l_1l_2)^2\delta^8(q_1^A+q_2^A).
\end{eqnarray}
Again the form of the integrand coincides with the $D=4$ case (see Fig.\ref{expan}).

Let us now briefly discuss the situation in $D=10$ dimensions.
The $D=10$ $\mathcal{N}=1$ SYM supermultiplet of the on-shell states consists of the 
physical polarizations of the gluon $A^{AB'}$ and fermion $\Psi^{A}$ fields.
In this case, one encounters the following difficulty in the attempt to construct the corresponding on-shell momentum
superspace: there are too many $\eta$ variables
\cite{SpinorHelisity_extraDimentions} to combine all the on-shell states in a manifestly Lorentz invariant manner. 
We need $4$ $\eta$ variables to accommodate all the on-shell $2^4$ states in the theory, but the smallest representation of the little group $SO(8)$ gives $8$  $\eta$'s. This problem, most likely, can be solved by using a modification of the harmonic superspace approach \cite{10dOnShell}, though the resulting structure of the tree level amplitude looks complicated and no unitarity based computations were performed so far in such a setup.

However, one can use the indirect symmetry arguments \cite{SpinorHelisity_extraDimentions} to show that the ratio of $\mathcal{A}_4^{(L)}/\mathcal{A}_4^{(0)}$ in $D=10$ $\mathcal{N}=1$ SYM has the form identical to that in the $D=4,6,8$ SYM theories(see also \cite{PureSpinorsMarfa}).

\section{The structure of UV divergences in the leading, subleading, etc. orders of PT in SYM theories}

In order to calculate the amplitude, it is convenient first to extract the color ordered partial amplitude by executing the color decomposition~\cite{Reviews_Ampl_General}
\begin{equation}
\mathcal{A}_n^{a_1\dots a_n,phys.}(p_1^{\lambda_1}\dots p_n^{\lambda_n})=\sum_{\sigma \in S_n/Z_n}Tr[\sigma(T^{a_1}\dots T^{a_n})]
\mathcal{A}_n(\sigma(p_1^{\lambda_1}\dots p_n^{\lambda_n}))+\mathcal{O}(1/N_c).
\end{equation}
The colour ordered amplitude $A_n$ is evaluated in the  limit  $N_c\to \infty$, $g^2_{YM}\to 0$ and $g^2_{YM}N_c$ is fixed, which corresponds to the planar diagrams.
In case of the four-point amplitudes, the colour decomposition is performed as follows:
\begin{eqnarray}
\mathcal{A}_4^{a_1\dots a_4,(L),phys.}(1,2,3,4)=T^1\mathcal{A}_4^{(L)}(1,2,3,4)+T^2\mathcal{A}_4^{(L)}(1,2,4,3)+
T^3\mathcal{A}_4^{(L)}(1,4,2,3)
\end{eqnarray}
where $T^i$ denote the trace combinations of $SU(N_c)$ generators in the fundamental representation 
\begin{gather}
T^1=Tr(T^{a_1}T^{a_2}T^{a_3}T^{a_4})+Tr(T^{a_1}T^{a_4}T^{a_3}T^{a_2}),\nonumber \\
T^2=Tr(T^{a_1}T^{a_2}T^{a_4}T^{a_3})+Tr(T^{a_1}T^{a_3}T^{a_4}T^{a_2}),\\
T^3=Tr(T^{a_1}T^{a_4}T^{a_2}T^{a_3})+Tr(T^{a_1}T^{a_3}T^{a_2}T^{a_4}).\nonumber
\end{gather}

\begin{figure}[htb]
 \begin{tabular}{c}
 $\frac{\mathcal{A}_4}{\mathcal{A}_4^{(0)}}=1+\sum\limits_L M^{(L)}_4(s,t)=$  \\
\includegraphics[scale=0.35]{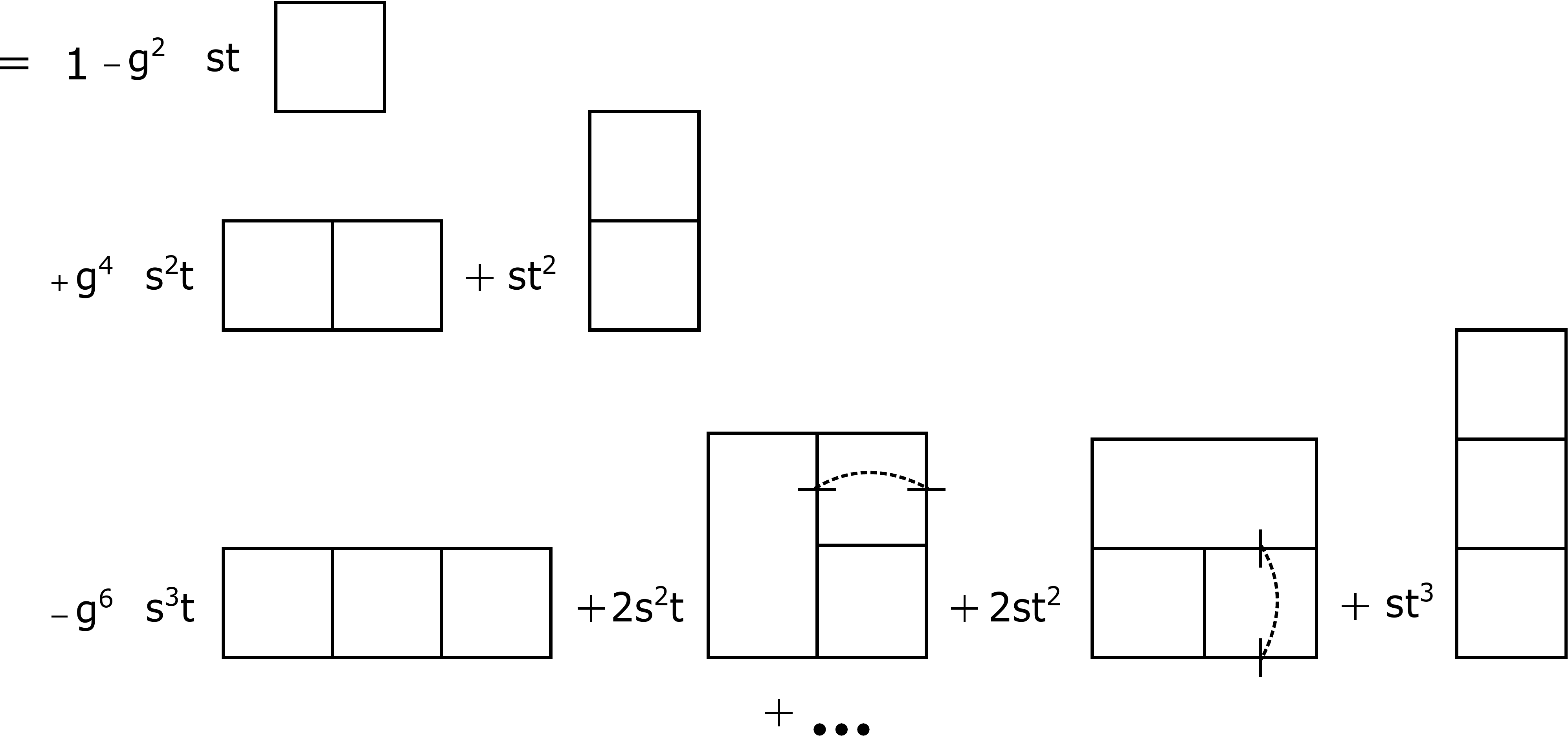}
\end{tabular}
 \caption{The universal expansion for the four-point scattering amplitude in SYM theories in terms of master integrals.
 The connected strokes on the lines mean  the square of the flowing momentum.}\label{expan}
 \end{figure}

The four-point tree-level amplitude is always factorized which is obvious within the superspace formalism. Hence the colour decomposed L-loop amplitude can be represented as
\beq
\mathcal{A}_4^{(L)}(1,2,3,4)=\mathcal{A}_4^{(0)}(1,2,3,4)M_4^{(L)}(1,2,3,4)=
\mathcal{A}_4^{(0)}(1,2,3,4)M_4^{(L)}([1+2]^2,[2+3]^2)\nonumber
\eeq
or using the standard Mandelstam variables
\beq
\mathcal{A}_4^{(L)}(1,2,3,4)=\mathcal{A}_4^{(0)}(1,2,3,4)M_4^{(L)}(s,t)
\eeq
The factorized amplitude $M_4^{(L)}(s,t)$ is the subject of calculation in this paper. Remarkably, it can be expressed in terms of some combination (which is universal for $D=6,8,10$ dimensions) of the pure scalar master integrals times some polynomial in the Mandelstam variables shown in Fig.\ref{expan}~\cite{Bern:2005iz}. 

Dimensional regularization (dimensional reduction) is common for calculation of UV divergences since the latter is manifested as pole terms with the numerators being the polynomials over the kinematic variables. 
In D-dimensions the first UV divergences start from L=6/(D-4) loops; therefore, in D=8 and D=10 SYM theories they start already at one loop. As one can see in Fig.\ref{expan},
all the bubble subgraphs as well as triangles are entirely omitted in PT of any order, which is the result of maximal supersymmetry, and appear only in less symmetric cases. In the D = 4 N = 4 case, this means the abandonment of all the UV divergences since the boxes are finite whereas at higher dimensions they are non-renormalizable by power counting. 

In general, $M_4^{(L)}$ has the form
\beq
M_4^{(L)}(s,t)=(-g^2)^L \sum_{i} \mbox{coef}_i \times \mbox{MasterIntegral}_i,
\eeq
where $g^2 \equiv \frac{g_{YM}^2N_c}{(4\pi)^{D/2}}$, the $\mbox{coef}_i$ are some monomials of $s$ and $t$, the $\mbox{MasterIntegral}_i$ is one of the master integrals in $D$-dimensional Minkowski space shown in Fig.\ref{expan}. The complete list of the master integrals up to 5 loops is presented in \cite{Bern:2005iz}.These master integrals are universal for any dimension. 

We use the following definition of the L-loop master integrals applied throughout the paper
\begin{eqnarray}\label{2}
\mbox{MasterIntegral}_i=\left(\frac{1}{i\pi^{D/2}}\right)^L\int d^Dk_1...d^Dk_L \frac{Num_i.}{Den_i.}.
\end{eqnarray}
Since we are interested in the UV divergences only, there is no need to calculate the multiloop diagram itself. The task is reduced to the extraction of the pole terms that essentially simplifies the calculation. To do this, we use the BPHZ $\R$-operation~\cite{BPHZ}.

For any local quantum field theory it is inherent that after performing the incomplete $\R$-operation, e.g $\R^\prime$-operation, the remaining UV divergences are always local. Due to this peculiarity it is possible to produce the so-called recurrence relations which link the divergent contributions in all orders of perturbation theory (PT) with the ones of the lower order. These relations are known as pole equations (within dimensional regularization) in renormalizable theories~\cite{hooft} and can be expressed in the form of the renormalization group. This holds true for any local theory, though can be trickier to execute technically, as we have shown in \cite{we2, we3}.  We recall the main steps of this procedure  below.

The incomplete ${\cal R}$-operation (${\cal R}^\prime$-operation) subtracts only the subdivergences of a given graph  while the full $\R$-operation is defined as 
\begin{equation}
\R G = (1-\K) \R' G, 
\end{equation}
where ${\cal K}$ is an operator that extracts out the singular part of the graph and $K{\cal R}^\prime G$- is the counter term corresponding to the graph G.
 Applying the ${\cal R}^\prime$-operation to a given graph $G$ in the n-th order of PT, one gets a series of terms presented below :
\beqa
{\cal R'}G_n&=&\frac{\A_n^{(n)}(\mu^2)^{n\epsilon}}{\epsilon^n}+\frac{\A_{n-1}^{(n)}(\mu^2)^{(n-1)\epsilon}}{\epsilon^n}+ ... +\frac{\A_1^{(n)}(\mu^2)^{\epsilon}}{\epsilon^n}\nonumber \\
&+&\frac{\B_n^{(n)}(\mu^2)^{n\epsilon}}{\epsilon^{n-1}}+\frac{\B_{n-1}^{(n)}(\mu^2)^{(n-1)\epsilon}}{\epsilon^{n-1}}+ ... +\frac{\B_1^{(n)}(\mu^2)^{\epsilon}}{\epsilon^{n-1}} \nonumber \\
&+&\frac{C_n^{(n)}(\mu^2)^{n\epsilon}}{\epsilon^{n-2}}+\frac{C_{n-1}^{(n)}(\mu^2)^{(n-1)\epsilon}}{\epsilon^{n-2}}+ ... +\frac{C_1^{(n)}(\mu^2)^{\epsilon}}{\epsilon^{n-2}} \nonumber\\
&+&\mbox{lower\ pole\ terms,}\label{Rn}
\eeqa
where the terms like $\frac{\A_{k}^{(n)}(\mu^2)^{k\epsilon}}{\epsilon^n}$  or $\frac{\B_{k}^{(n)}(\mu^2)^{k\epsilon}}{\epsilon^{n-1}}$, $\frac{C_{k}^{(n)}(\mu^2)^{k\epsilon}}{\epsilon^{n-2}}$ originate from the $k$-loop graph which remains after subtraction of the $(n-k)$-loop counterterm.
The resulting expression has to be local and hence does not contain terms like $\log^l{\mu^2}/\epsilon^k$ from any $l$ and $k$. This requirement leads to a sequence of relations for $\A_i^{(n)}, \B_i^{(n)}$ and $C_i^{(n)}$ which can be solved in favour of the lowest order terms
\beqa
\A_n^{(n)}&=&(-1)^{n+1}\frac{\A_1^{(n)}}{n}, \nonumber\\
\B_n^{(n)}&=&(-1)^n \left(\frac 2n \B_2^{(n)}+\frac{n-2}{n}\B_1^{(n)}\right) \label{rel},\\
C_n^{(n)}&=&(-1)^{n+1}\left(\frac{3}{n}C_3^{(n)}+\frac{2(n-3)}{n}C_2^{(n)}+\frac{(n-2)(n-3)}{2n}C_1^{(n)}\right).\nonumber \label{abc_coeff}
\eeqa
It is also useful to write down the local expression for the ${\cal KR'}$ terms (counterterms) equal to
\beq
{\cal KR'}G_n=\sum_{k=1}^n \left(\frac{\A_k^{(n)}}{\epsilon^n} +\frac{\B_k^{(n)}}{\epsilon^{n-1}}+\frac{C_k^{(n)}}{\epsilon^{n-2}}+\cdots \right)\equiv
\frac{\A_n^{(n)'}}{\epsilon^n}+\frac{\B_n^{(n)'}}{\epsilon^{n-1}}+\frac{C_n^{(n)'}}{\epsilon^{n-2}}+ \cdots.
\eeq
Then, one has, respectively,
\beqa
\A_n^{(n)'}&=&(-1)^{n+1}\A_n^{(n)}=\frac{\A_1^{(n)}}{n}, \nonumber \\
\B_n^{(n)'}&=& \left(\frac{2}{n(n-1)} \B_2^{(n)}+\frac{2}{n}\B_1^{(n)}\right) \label{rel2},\\
C_n^{(n)'}&=&\left(\frac{2}{(n-1)(n-2)}\frac{3}{n}C_3^{(n)}+\frac{2}{n-1}\frac{3}{n}C_2^{(n)}+\frac{3}{n}C_1^{(n)}\right). 
\nonumber
\eeqa
	
This means that performing the  ${\cal R}'$-operation of the higher order diagrams, it is possible to  deal only with the one-, two-, or three-loop subgraphs surviving after contraction of subdivergences and get the desired leading pole terms via eqs.(\ref{rel})  in the leading, subleading and sub-subleading order, respectively. The latter can be evaluated in all loops algebraically.

\begin{figure}[htb!]
\begin{center}
\leavevmode
\includegraphics[width=0.8\textwidth]{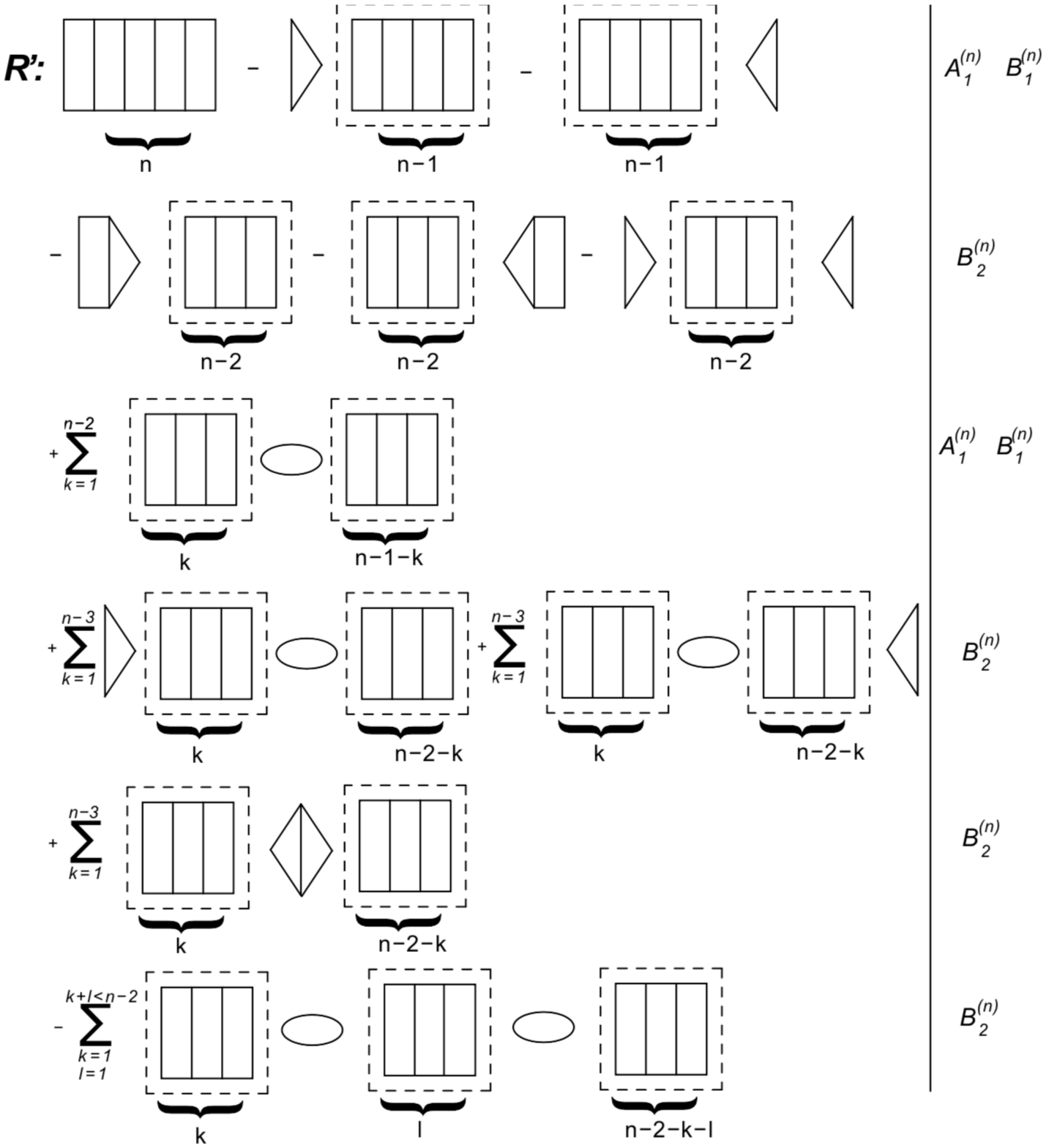}
\end{center}\vspace{-1cm}
\caption{The $\R'$-operation for the horizontal ladder in D=8}
\label{R}
\end{figure}

The discussed procedure makes it possible not only to derive solutions for a fixed number of loops but also to obtain the recurrence relations in any loop order. We demonstrate this derivation by the example of the horizontal ladder-type diagrams in $D=8$~\cite{we3} (see Fig.\ref{R}). 

We start with the leading order. First, one can simplify the notation $ A_n^{(n)}=s^{n-1}A_n$ and $ A_n^{(n)'}=s^{n-1}A'_n$ since the horizontal ladder-type diagrams in the leading order depend only on s. By calculating the one-loop diagrams shown in the first and third lines of Fig.\ref{R} and substituting them into eq.(\ref{rel}), we receive the recurrence relation in the leading order.
\beq
n A_n=-\frac{2}{4!} A_{n-1}+\frac{2}{5!}\sum_{k=1}^{n-2}A_kA_{n-1-k}, \ \ \  n\geq 2, \label{one8}
\eeq
where $A_1=1/3!$. Starting from one-loop term, the leading divergence in any loop order can be calculated using this recurrence relation solely algebraically.

In the subleading order one has already the t dependence but it is linear. To separate it,  we use the notation
 $B_n^{(n)}=s^{n-1}B_{ns-1}+s^{n-2}tB_{tn}$ and $B_n^{(n)'}=s^{n-1}B^{'}_{ns-1}+s^{n-2}tB^{'}_{tn}$. 
In order to derive the recurrence relation in the subleading case, one has to calculate the two-loop diagrams shown in the second and last lines of Fig.\ref{R}. We begin with the primed quantities since they actually enter into the recurrence relations

\newpage
 
 \beqa
B'_{tn}&=&-\frac{2}{n(n-1)}B'_{tn-2}\frac{10}{5!5!}+\frac 2n B'_{tn-1}\frac{2}{5!}, \label{btprime}
 \eeqa
  \beqa
B'_{sn}&=&\frac{2}{n(n-1)}\left[-A'_{n-2}\frac{2321}{5!5!2}-B'_{sn-2}\frac{18}{4!5!}+B'_{tn-2}\frac{44}{5!5!}\right. \nonumber\\
&-& \left. \sum_{k=1}^{n-3}A'_kA'_{n-2-k}\frac{938}{4!5!15}- \sum_{k=1}^{n-3}A'_kB'_{sn-2-k}\frac{1}{5!2}+
 \sum_{k=1}^{n-3}A'_kB'_{tn-2-k}\frac{442}{5!5!12}\right. \nonumber\\
 &-&\left. \sum_{k,l=1}^{n-k+l<n-2}A'_kA'_lA'_{n-2-k-l}\frac{8}{5!5!}\frac{46}{15}
  -\sum_{k,l=1}^{n-k+l<n-2}A'_kA'_lB'_{sn-2-k-l}\frac{12}{5!5!}\right. \nonumber \\
 &+&\left. \sum_{k,l=1}^{n-k+l<n-2}A'_kA'_lB'_{tn-2-k-l}\frac{4}{5!5!}
  +\sum_{k,l=1}^{n-k+l<n-2}B'_kA'_lA'_{sn-2-k-l}\frac{2}{5!5!}\right]\nonumber \\
  &+&\frac 2n\left[ A'_{n-1}\frac{19}{3 4!}+B'_{sn-1}\frac{2}{4!}-B'_{tn-1}\frac{4}{5!}\right. \nonumber \\
 &+& \left. \sum_{k=1}^{n-2}A'_kA'_{n-1-k}\frac{2}{5!}\frac{46}{15}+\sum_{k=1}^{n-2}A'_kB'_{sn-1-k}\frac{4}{5!}-
 \sum_{k=1}^{n-2}A'_kB'_{tn-1-k}\frac{2}{5!}\right] .\label{bsprime}
\eeqa
where $B'_{s1}=B'_{t1}=0$, $B'_{s2}=-5/3!/4!/12$, $B'_{t2}=-1/3!/4!/6$. 

Proceeding in a similar way one can get relations for the unprimed quantities. The recurrence relations for the sub-subleading divergences are not presented here due to their length. 

Solution of the recurrence relations (\ref{one8},\ref{btprime},\ref{bsprime}) is complicated. However, since we actually need the sum of the series, we perform the summation
multiplying both sides of eq.(\ref{one8}) by $z^{n-1}$ and take the sum from 3 to infinity. After some algebraic manipulation and introducing the notation  $\Sigma_A=\sum_{n=1}^\infty A_n (-z)^n$,  we
finally transform the recurrence relation to the differential equation. In the leading order we get (here $z \equiv g^2 s^2/\epsilon$)
\beq
\frac{d}{dz}\Sigma_A=-\frac{1}{3!}+\frac{2}{4!}\Sigma_A-\frac{2}{5!}\Sigma_A^2. \label{eqa}
\eeq

Similar differential  equations can be constructed for $\Sigma'_{sB}=\sum_2^\infty z^nB'_{sn}$ and $\Sigma'_{tB}=\sum_2^\infty z^nB'_{tn}$,
\beq
\frac{d^2 \Sigma'_{tB}(z)}{dz^2}-\frac{1}{30}\frac{d \Sigma'_{tB}(z)}{dz}+\frac{\Sigma'_{tB}(z)}{720}=-\frac{1}{432},
\label{eq1}
\eeq
\beq
\frac{d^2 \Sigma'_{sB}(z)}{dz^2}+f_1(z)\frac{d \Sigma'_{sB}(z)}{dz}+f_2(z)\Sigma'_{sB}(z)=f_3(z), \label{Ric}
\eeq
with
\beqa
f_1(z)&=&-\frac 16+\frac{\Sigma_A}{15},\nonumber\\
f_2(z)&=&\frac{1}{80}-\frac{\Sigma_A}{120}+\frac{\Sigma_A^2}{600}+\frac{1}{15}\frac{d \Sigma_A}{dz}, \nonumber\\
f_3(z)&=&\frac{2321}{5!5!2}\Sigma_A+\frac{11}{1800}\Sigma'_{tB}-\frac{469}{5!90}\Sigma_A^2-\frac{442}{5!5!6}\Sigma_A\Sigma'_{tB}+\frac{23}{6750}\Sigma_A^3+\frac{1}{1200}\Sigma_A^2\Sigma'_{tB}\nonumber\\
&-&\frac{19}{36}\frac{d \Sigma_A}{dz}-\frac{1}{15}\frac{d \Sigma'_{tB}}{dz}
+\frac{23}{225}\frac{d \Sigma_A^2}{dz}+\frac{1}{30}\frac{d( \Sigma_A\Sigma'_{tB})}{dz}-\frac{3}{32}.\nonumber
\eeqa

One can perform the same procedure for a specific series of diagrams as well as for the entire set by using some symmetry arguments.
In \cite{we2}, we constructed the full recurrence relations for the leading divergences for SYM theories in D = 6 and D = 8, 10. This was performed by consistent application of the $\R'$-operation and integration over the remaining triangle and bubble diagrams with the help of Feynman parameters. While executing this we notice that the full set of UV divergent diagrams (master integrals) consists of the s-channel and t-channel ones, and it is needed simply to add the box to the corresponding channel in order to move from $n-1$ to $n$ loops.
Denoting by  $S_n(s,t)$ and  $T_n(s,t)$  the sum of all contributions in the  $n$-th order of PT in $s$ and  $t$ channels, respectively, so that
\beq
\frac{\mathcal{A}_4}{\mathcal{A}_4^{(0)}}\bigg|_{\mbox{leading UV div.}}=\sum_{n=0}^{\infty}g^{2n}\frac{S_n(s,t)+T_n(s,t)}{\epsilon^n},
\eeq
we get the following recursive relations for the $D=6,8$ and $D=10$ cases, respectively:
\beq
nS_n(s,t)=-2 s \int_0^1 dx \int_0^x dy  \ (S_{n-1}(s,t')+T_{n-1}(s,t')), \ \ \ \     n\geq 4\label{req6}
\eeq
where $t'= t x+u y$, $u=-t-s$, and   $S_3=-s/3,\ T_3=-t/3$.
\beqa
&&nS_n(s,t)=-2 s^2 \int_0^1 dx \int_0^x dy\  y(1-x) \ (S_{n-1}(s,t')+T_{n-1}(s,t'))|_{t'=tx+uy}\nonumber \\ &+&
s^4 \int_0^1\! dx \ x^2(1-x)^2 \sum_{k=1}^{n-2}  \sum_{p=0}^{2k-2} \frac{1}{p!(p+2)!} \
 \frac{d^p}{dt'^p}(S_{k}(s,t')+T_{k}(s,t')) \times \nonumber \\
&&\hspace{2cm}\times  \frac{d^p}{dt'^p}(S_{n-1-k}(s,t')+T_{n-1-k}(s,t'))|_{t'=-sx} \ (tsx(1-x))^p, \label{req8}
\eeqa
where $S_1= \frac{1}{12},\ T_1=\frac{1}{12}$.
\beqa
&&nS_n(s,t)=-s^3 \int_0^1 dx \int_0^x dy\  y^2(1-x)^2 \ (S_{n-1}(s,t')+T_{n-1}(s,t'))|_{t'=tx+yu}\label{req10}\nonumber \\ &+&
s^5 \int_0^1\! dx \ x^3(1-x)^3 \sum_{k=1}^{n-2}  \sum_{p=0}^{3k-2} \frac{1}{p!(p+3)!} \
 \frac{d^p}{dt'^p}(S_{k}(s,t')+T_{k}(s,t')) \times \nonumber \\
&&\hspace{2cm}\times  \frac{d^p}{dt'^p}(S_{n-1-k}(s,t')+T_{n-1-k}(s,t'))|_{t'=-sx} \ (tsx(1-x))^p, 
\eeqa
where $S_1= \frac{s}{5!},\ T_1=\frac{t}{5!}$.
The leading divergences in any order of PT can be designed in algebraic form using these recurrence relations, starting from the known values of $S_1$ and $T_1$.

Similar to the ladder case, these recurrence relations include all the diagrams of a given order of PT and allow to sum all orders of PT. This can be done by multiplying both sides of eqs.(\ref{req6},\ref{req8},\ref{req10}) by $(-z)^{n-1}$, where $z=\frac{g^2}{\epsilon}$ and summing up from n=2 to infinity. Denoting the sum by $\Sigma(s,t,z)=\sum_{n=1}^\infty S_n(s,t) (-z)^n$, we finally obtain the following differential equations
in the $D=6,8$ and $D=10$ cases:
\beq
 \frac{d}{dz}\Sigma(s,t,z)=s-\frac{2}{z}\Sigma(s,t,z)+2s \int_0^1 dx \int_0^x dy\ (\Sigma(s,t',z)+\Sigma(t',s,z))|_{t'=xt+yu}.
\label{eq6}
\eeq
\beqa
&&\frac{d}{dz}\Sigma(s,t,z)=-\frac{1}{12}+2 s^2 \int_0^1 dx \int_0^x dy\  y(1-x)\ (\Sigma(s,t',z)+\Sigma(t',s,z))|_{t'=tx+uy}
\label{eq8}\\
&&-s^4  \int_0^1\! dx \ x^2(1-x)^2 \sum_{p=0}^\infty \frac{1}{p!(p+2)!} (\frac{d^p}{dt'^p}(\Sigma(s,t',z)+\Sigma(t',s,z))|_{t'=-sx})^2 \ (tsx(1-x))^p. \nonumber
\eeqa
\beqa
&&\frac{d}{dz}\Sigma(s,t,z)=-\frac{s}{5!}+s^3 \int_0^1 dx \int_0^x dy\  y^2(1-x)^2\ (\Sigma(s,t',z)+\Sigma(t',s,z))|_{t'=tx+yu}\\
&&-s^5  \int_0^1\! dx \ x^3(1-x)^3 \sum_{p=0}^\infty \frac{1}{p!(p+3)!} (\frac{d^p}{dt'^p}(\Sigma(s,t',z)+\Sigma(t',s,z))|_{t'=-sx})^2 \ (tsx(1-x))^p.
\nonumber
\label{eq10}
\eeqa
The same equations with the replacement $s \leftrightarrow t$ are valid for $\Sigma(t,s,z)=\sum_{n=1}^\infty T_n(s,t) (-z)^n$.

\section{Properties of the solutions and numerical analysis}

Since eqs.(\ref{eq6} - \ref{eq10}) are integro-differential, their analytical solution is problematic. Therefore, we use the ladder type diagrams, which are much simpler and allow for the explicit solution, as an approximation to the solution of the exact equations. We show that the ladder type diagrams are in good agreement with the total PT series and may serve as a model for the full answer.

\subsection{The Ladder case}

{\bf D=6}

\begin{figure}[htb!]
\begin{center}
\leavevmode
\includegraphics[width=0.8\textwidth]{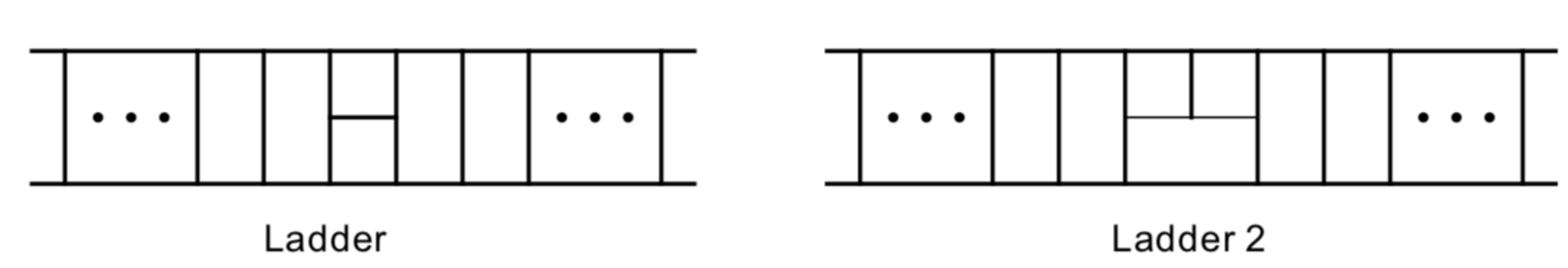}
\end{center}
\caption{The ladder type diagrams in D=6}
\label{laddiag}
\end{figure}

The D=6 case is of particular interest since the boxes are finite. Therefore, the s-ladder type diagram of interest contains one tennis-court subdiagram and the ladder added from the left or right (see Fig.\ref{laddiag}, left).

Using the recursive relations (see \cite{we2}), one can obtain the equation for the ladder diagrams; however, it is also possible to derive it from eq.(\ref{eq6}). These diagrams contain only $s$ - dependence, so it drops out from the integrals in the r.h.s. of eq.(\ref{eq6}) for the first term. The second term corresponds to the t-ladder subdiagrams and does not give a contribution to the ladder approximation. As a result, we have the ordinary differential equation
\beq
\frac{d \Sigma_L(s,z)}{dz}=s-\frac 2z \Sigma_L(s,z)+s\Sigma_L(s,z)=0, \ \ \ \ \Sigma_L(s,0)=0.
\eeq
Here $\Sigma_L(s,z)$ is dimensionless and depends on a single dimensionless argument $sz$.

Solution to this equation is
\beq
\Sigma_L(s,z)=\frac{2}{s^2z^2}(e^{sz}-1-sz-\frac{s^2z^2}{2}).\label{lad6}
\eeq
And for the vertical ladder we have the same solution with the replacement $s\leftrightarrow t$.

Depending on the sign of $s$, the obtained solution tends either to infinity or to a constant when $z\to\infty\ (\epsilon\to 0)$. We show further that the full solution has the same behaviour and discuss its consequences below.

One can derive a similar equation for the next sequence of ladder type diagrams, which starts from four loops (Fig.\ref{laddiag} right). The resulting expression also contains only $s$-dependence except for one power of $t$. This leads to two coupled recursive relations and hence two coupled differential equations. The solution of these equations has the form
\beqa
\Sigma_{L2}(s,t,z)&=&\frac{1}{2s^2z^2}\left[27(e^{sz/3}-1-\frac{sz}{3}-\frac 12\frac{s^2z^2}{9}-\frac 16\frac{s^3z^3}{27})(1+2\frac ts)\right.\nonumber\\
&&\left.-(e^{sz}-1-sz-\frac 12 s^2z^2-\frac 16 s^3z^3)\right].\label{lad62}
\eeqa
Depending on the sign of $s$, this solution has the same behaviour as the previous one, i.e. tends either to infinity or to a constant.
We see later that the sum of two ladders  (\ref{lad6}) and (\ref{lad62}) gives a better approximation to the solution of the full equation.\\

{\bf D=8}

In this case, the ladder starts already from one loop. It also contains only $s$ - dependence, so all the integrals in eq.(\ref{eq8}) are trivial for the first terms in the bracket while the second terms have no contributions to the s-ladder like in the previous case. Then eq.(\ref{eq8}) reduces to the ordinary nonlinear differential equation
\beq
\frac{d \Sigma_L(s,z)}{dz}=-\frac{1}{3!}+\frac{2}{4!} \Sigma_L(s,z)-\frac{2}{5!}\Sigma_L^2(s,z)=0, \ \ \ \ \Sigma_L(s,0)=0.
\eeq
Here $\Sigma_L(s,z)$ is also dimensionless and depends on the single dimensionless argument $s^2z$.

This equation refers to Riccati type equations with constant coefficients. The solution has the form
\beq
\Sigma_L(s,z)=-\sqrt{5/3} \frac{4 \tan(zs^2/(8 \sqrt{15}))}{1 - \tan(zs^2/(8 \sqrt{15}))\sqrt{5/3}}.\label{lad8}
\eeq

This function has an infinite number of periodical poles and there is no simple limit when $z\to\infty\ (\epsilon\to 0)$ regardless of kinematics. Further we will see that this property also characterizes the full solution.\\

{\bf D=10}

This case looks similar D=8 but becomes more complicated due to the genuine box diagram in D=10. Contrary to D=8, this diagram is not a constant but is proportional to $(s+t)$. Consequently, the s-ladder has dimension $m^2$ and consists of two parts, one proportional to s and the other to t times dimensionless function of $s^3z$
$$\Sigma_L(s,t,z)=s\Sigma_s(s,z)+t\Sigma_t(s,z).$$

Equation (\ref{eq10}) reduces to the ordinary nonlinear differential equation as in the D=8 case; however, we have two coupled equations for  $\Sigma_s(s,z)$ and $\Sigma_t(s,z)$. To obtain these equations in a simple way, one needs to use the recursive relations~\cite{we2}
\beqa
\frac{d \Sigma_t(s,z)}{dz}&=&
-\frac{1}{5!} +\frac{4}{7!}\Sigma_t(s,z) - \frac{1}{3*7!}\Sigma_t^2(s,z), \ \ \ \ \ \ \ \ \  \ \ \ \ \ \ \ \ \ \Sigma_t(s,0)=0,\\
\frac{d \Sigma_s(s,z)}{dz}&=& - \frac{1}{5!} +\frac{2}{3*5!}\Sigma_s(s,z) - \frac{12}{7!}\Sigma_t(s,z)\nonumber \\&-&\frac{3!}{7!}\left(\Sigma_s^2(s,z)-\Sigma_s(s,z)\Sigma_t(s,z)+\frac{5}{18}\Sigma_t^2(s,z)\right), \ \   \Sigma_s(s,0)=0.
\eeqa
Note that both functions are dimensionless and depend on the single dimensionless argument $s^3z$.

The solution of the first equation is
\beq
\Sigma_t(s,z)=3 \left(2 + \sqrt{10}\tan\left[\frac{-\sqrt{10} zs^3 - 5040 \arctan[\sqrt{2/5}]}{5040}\right]\right)\label{lad10}
\eeq
while the second one is expressed in the form
\beq
\Sigma_s(s,z) = \frac{1}{2}\Sigma_t(s,z) + \Delta(s,z)
\eeq
where the function $\Delta(s,z)$ is the solution to the nonlinear differential equation
\beq
\frac{d \Delta(s,z)}{dz}=
- \frac{1}{2*5!} +\frac{2}{3*5!}\Delta(s,z) - \frac{6}{7!}\Delta^2(s,z) = 0, \ \ \Delta(s,0)=0.\label{dd}
\eeq
This is also a dimensionless function of the single dimensionless variable $s^3z$. The solution of eq.(\ref{dd}) is
\beq
\Delta(s,z)=-\frac{(3 (14 + \sqrt{70}) (-1 + e^{zs^3/(36 \sqrt{70})})}{2 (19 + 2 \sqrt{70} - 9 e^{zs^3/(36 \sqrt{70})})}\label{ladd}
\eeq

The behaviour of $\Sigma_t$ is similar to $\Sigma_L$ in the D=8, i.e. it possesses an infinite number of periodical poles. There is also a single pole in the function $\Delta$ for positive values of $s$.

\subsection{The General Case}

In this subsection, we analyze equations  (\ref{eq6},\ref{eq8},\ref{eq10}) that give us the sum of infinite series of diagrams. 
Due to complexity of these equations, a numerical solution can be a suitable method, although this approach also has its difficulties.
We cannot use the standard recursive algorithm because unknown functions are under the integral and depend on integration variables in a complex way.

We use an algorithm that is a combination of the standard numerical method and the method of successive approximations. 
First of all, we select some initial value of the function $\Sigma(s,t,z)=\Sigma_0(s,t,z) = const$  and then start with it. If we start with $z_0=0$, then a suitable choice is $const=0$. 
After that we substitute it in the r.h.s. and perform formal integration $(\Sigma_1(s,t,z)-\Sigma_0(s,t,z))/\Delta z$ to get the following approximation for $\Sigma$:
\beq
\Sigma_1(s,t,z)=\Sigma_0(s,t,z)+\Delta z * r.h.s,  \eeq
which is now a polynomial over $s$ and $t$. At this step the r.h.s. is calculated with $\Sigma_0(s,t',z)$ equal to a constant.

The next step is setting up of the polynomial in the r.h.s.   Changing the arguments $t\rightarrow tx+uy$ and $t \rightarrow -sx$, we perform the integration. This generates the next approximation value of $\Sigma$: $\Sigma_2(s,t,z)$. Continuing this procedure, we generate the highest degree polynomials $s$ and $t$ at each step. However, starting with 3-4 iterations, the length of the  polynomials becomes too time consuming for further calculation. 
At this step we estimate the value of $\Sigma$ with the fixed values of s and t, for example, s = t = 1. The calculated value gives us a constant, which we identify with the value of $ \Sigma $ at $ z_0 + \Delta z $. 
We use this value to run the same procedure again for the next iteration. This way we calculate the values of $\Sigma$ at the points along the axis $z=z_0+ \Delta z * n$.

Then we interpolate all obtained points to a smooth function. We found that $\Delta z = 0.1$ makes the solutions stable. The numerical results demonstrate a reasonable approximation being applied to the  known functions, though this method is not justified. 

Note that after calculating the function $\Sigma$, we can replace its argument having in mind  that it depends on the dimensionless combinations $ zs, zs ^ 2 $ and $ zs ^ 3 $ (and the same for $ t $) for $ D = $ 6, 8 and 10, respectively, for dimensional reasons.

It should also be said that in the $ D =$ 8 and 10 cases the form of equations (\ref{eq8}, \ref{eq10}) is not suitable for numerical analysis since the second term contains an infinite sum with an infinite number of derivatives. Cutting this sum makes the numerical solution unstable. To avoid this problem, we note that the construction resembles an ordinary shift operator with slightly modified coefficients.  This infinite sum can be removed by introducing two additional integrations which do not cause difficulties in numerical integration. One has:
\beqa
&&\sum_{p=0}^\infty \frac{(BC)^p k!}{p!(p+k+1)!} \left(\frac{d^p}{dA^p}f(A)\right)^2=\\=
&&\frac{1}{2\pi}\int_{-\pi}^{\pi} d\tau \int_0^1 d\xi(1-\xi)^k f(A+exp(i\tau)B\xi) f(A+exp(-i\tau)C). \nonumber
\eeqa
 We use this technique for numerical calculations in the case of $ D = 8 $ and $ D = 10 $.

The results of application of the  described techniques  for all three cases $D=6,8,10$  are presented below.

To test our numerical procedure, we compare the results of our calculation with the results obtained using PT, the Pade approximation and the Ladder approximation.  For comparison, we use the first 15 terms of the PT series that are generated using our recurrence relations. This seems to be far enough since the successive PT coefficients are falling rapidly.

The next step is to use the Pade approximation. This is not always stable since Pade approximations sometimes have fictitious poles. It is a well-known feature, and we tried to avoid it using mainly diagonal approximations. With 15 terms of PT the (6,6), (6,7) and (7,7) approximants are almost identical and give a smooth function. 

The third curve in the graphs corresponds to the ladder approximation. Analytical solutions here are given by eqs.(\ref {lad6}, \ref {lad8}, \ref {lad10}, \ref {ladd}) from the previous subsection. In the case of D = 6, we also considered the second ladder, which is based on the tennis court diagram in the t-channel (see Fig. \ref {laddiag}) and is given by eq.(\ref {lad62}).

Finally, we build a numerical solution obtained by the iteration procedure described above. In the case when the function has poles, we build a numerical solution separately for each finite interval.

The function $ \Sigma (s, t, z)$  is a function of three variables. However, as was already mentioned, for dimensional reasons, it has only two independent dimensionless arguments. In $ D =$ 6, 8 and 10 they   are $ zs, zt $, $ zs ^ 2, zt ^ 2 $ and $ zs ^ 3, zt ^ 3 $, respectively. We constructed both two-dimensional graphs with $ t = s $ and three-dimensional graphs in the $ s-t $ plane  for  better presentation.\\

{\bf D=6}

In D=6 the PT series looks like
\beqa
\Sigma_{PT}(s,t,z)&=& \frac{(s + t)z}{3} + \frac{(s^2 + st + t^2)z^2}{18} + \frac{(5s^3 + 2s^2t + 2st^2 + 5t^3)z^3}{540}+
\nonumber\\
&+&\frac{ (25 s^4 + 8 s^3 t - 2 s^2 t^2 + 8 s t^3 + 25 t^4) z^4}{19440}  + ... ,\label{pt6}
\eeqa
where the dots stand for the higher order terms. We used 15 terms for numerical comparison with the other approaches.

From eq.(\ref{pt6}) we constructed the diagonal Pade approximant [7/7]  as a function of a new variable $x=zs$ in the case when $t=s$. It has the form
\beqa
\Sigma_{Pade}(x)= && \frac{0.67 x + 0.067 x^2 + 0.0010 x^3 +  0.00014 x^4 + 4.6\cdot10^{-5} x^5 +}{1 - 0.15 x + 0.00013 x^2  +
 0.0011 x^3 - 4.5\cdot10^{-5} x^4 - 2.1\cdot10^{-6} x^5 +} \rightarrow  \nonumber \\
&& \leftarrow \frac{ + 3.7\cdot10^{-6} x^6 + 1.2\cdot10^{-7} x^7}{+ 1.7\cdot10^{-7} x^6 -2.1\cdot10^{-9} x^7}
 \label{pade6}
\eeqa

\begin{figure}[htb!]
\begin{center}
\leavevmode
\includegraphics[width=0.5\textwidth]{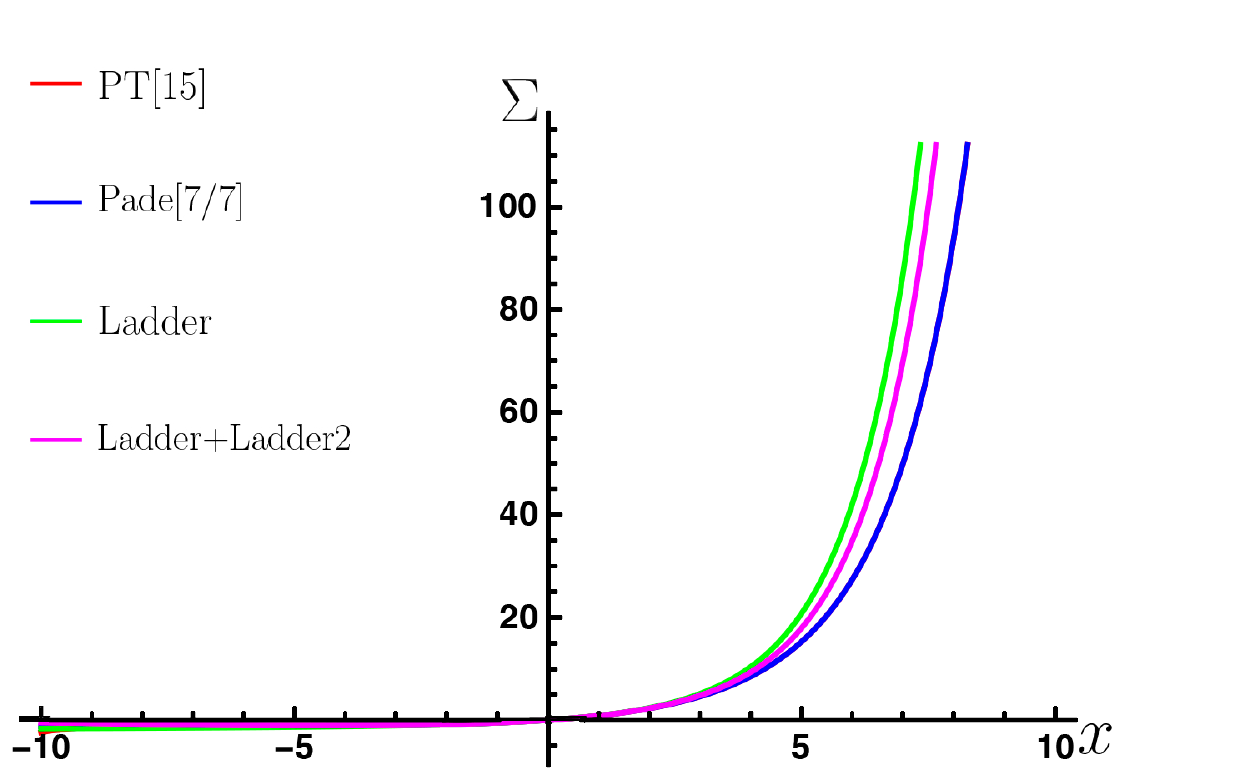}
\end{center}
\caption{Comparison of various approaches to solve eq.(\ref{eq6}): PT, Pade, Ladder and Numerics. The PT curve and the Pade one coincide in a given interval.}

\label{allloop6}
\end{figure}

\begin{figure}[htb!]
\begin{center}
\leavevmode
\includegraphics[width=0.3\textwidth]{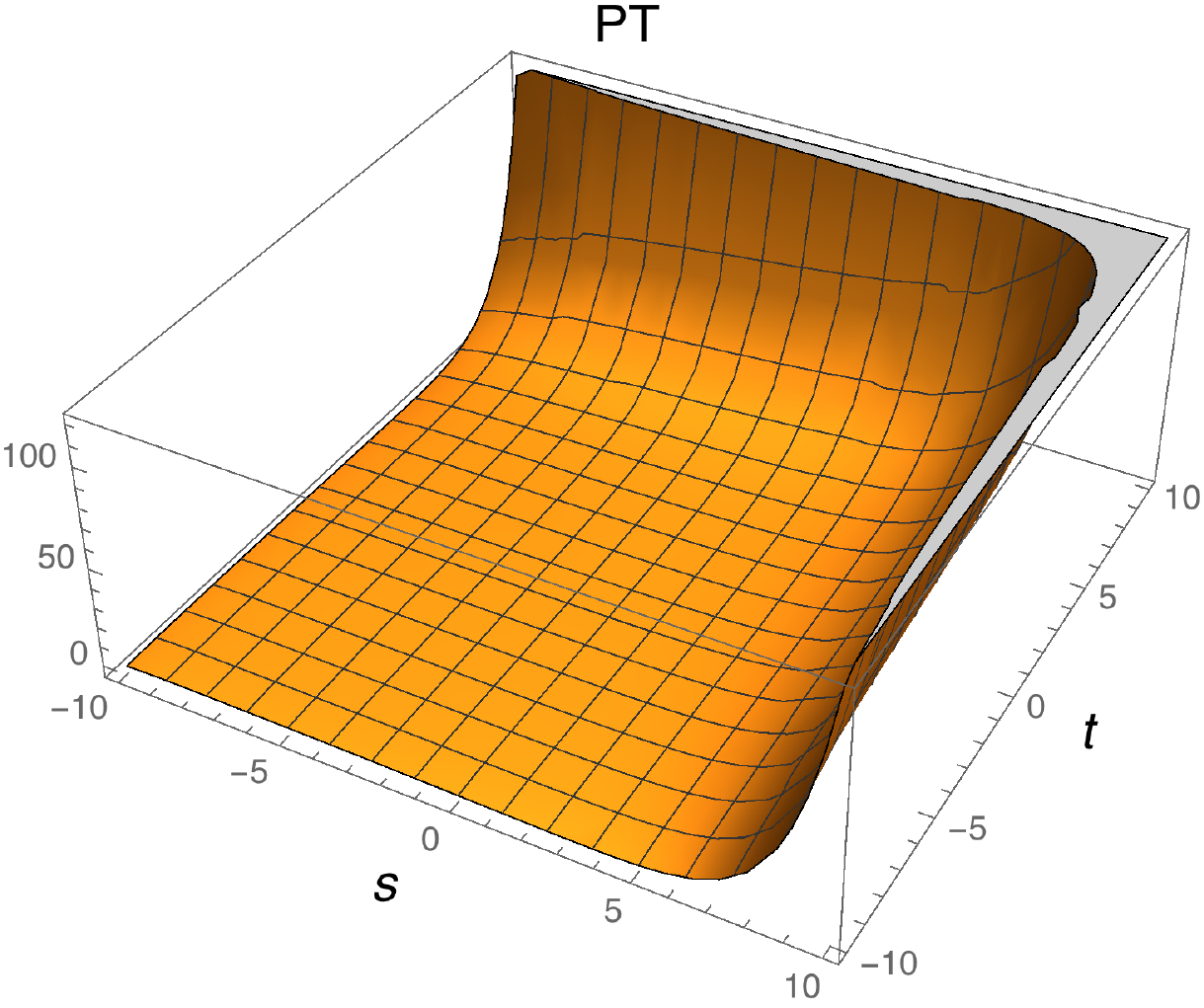}
\includegraphics[width=0.3\textwidth]{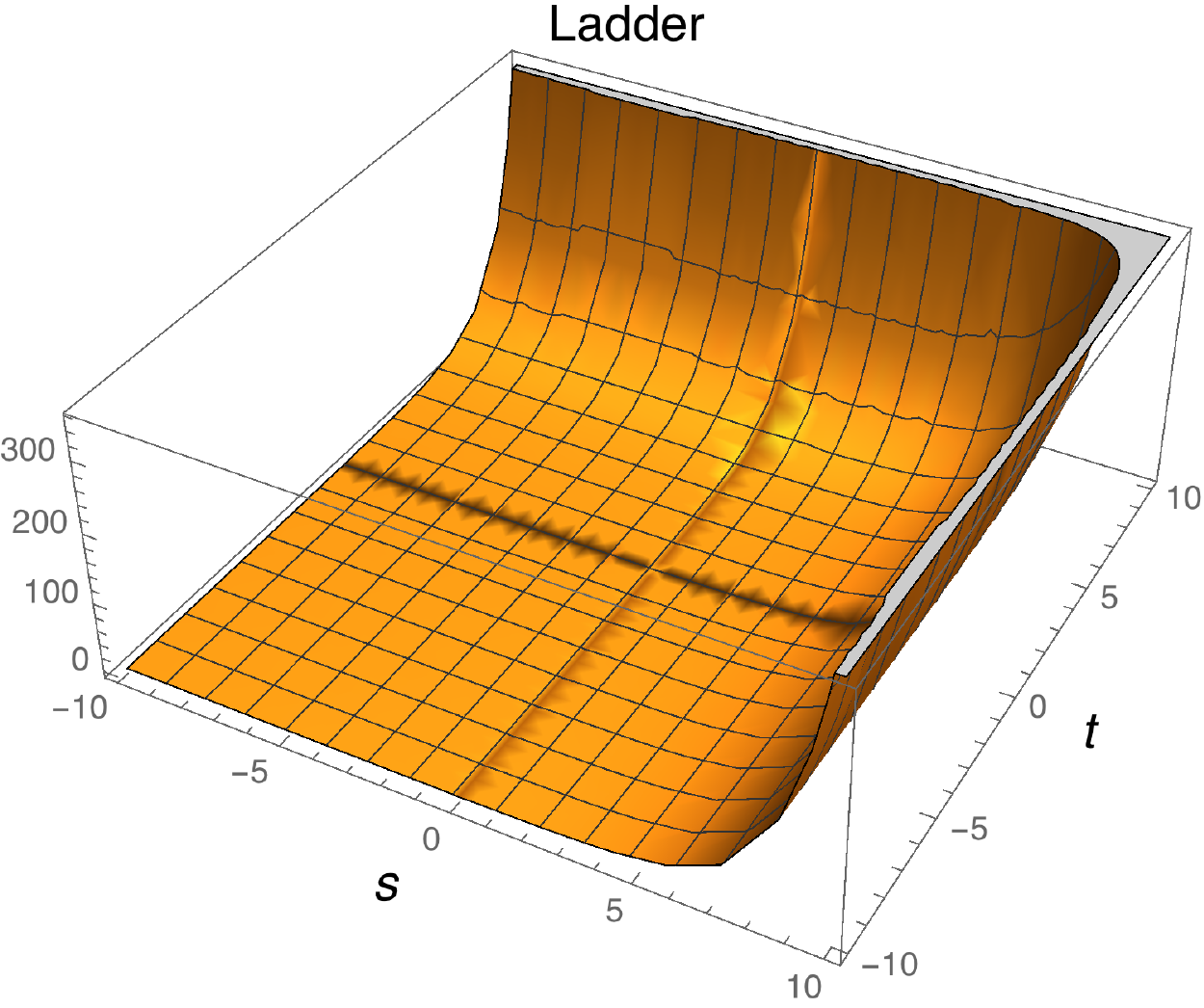}
\includegraphics[width=0.3\textwidth]{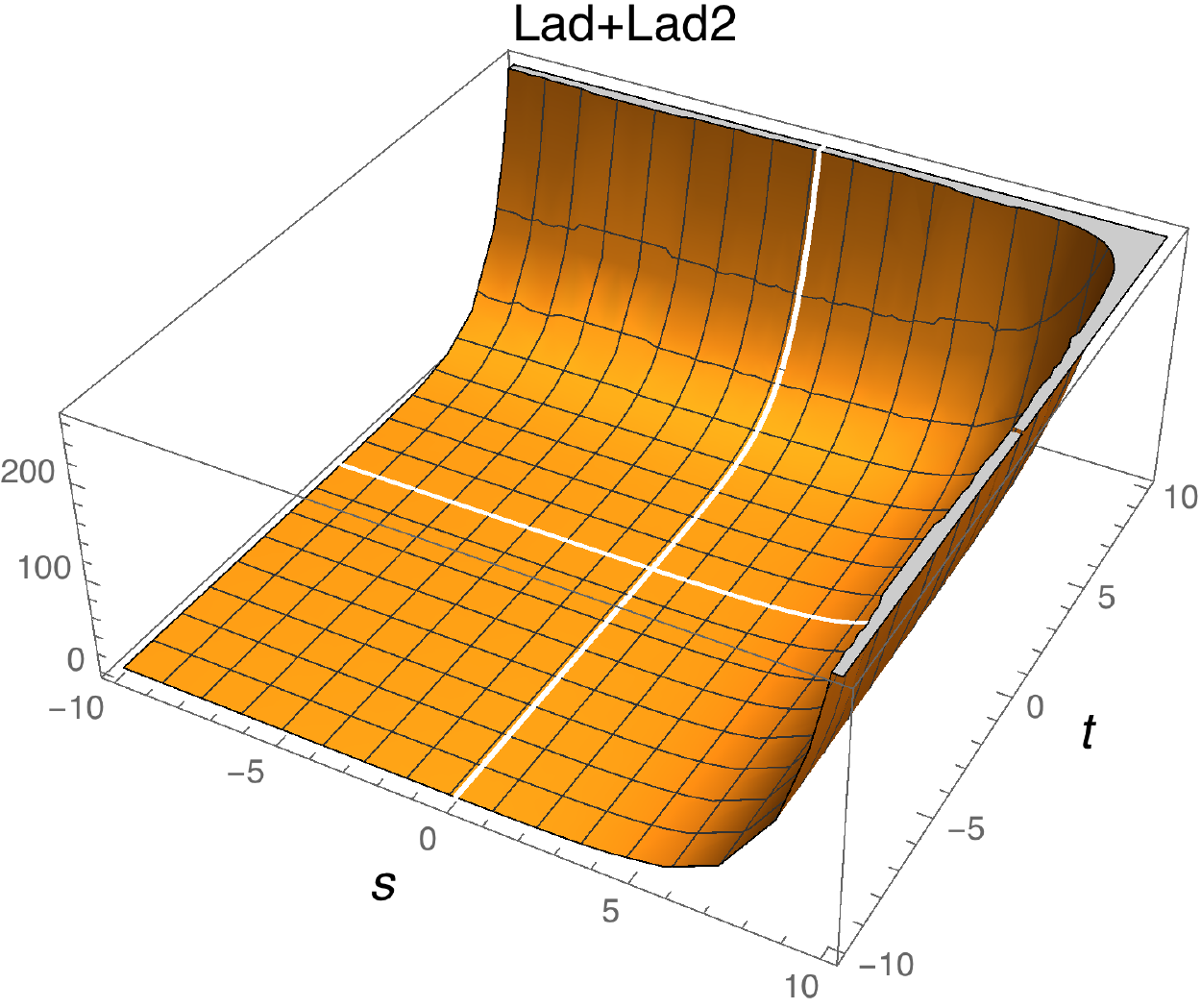}
\end{center}
\caption{Comparison of  PT, Ladder and Ladder2.}
\label{3d6}
\end{figure}

The ladder approximation is given by eq.(\ref{lad6}) and the second ladder by eq.(\ref{lad62}) with $x=zs$.
The numerical solution starts from $z=0$ and  in this case has only one interval.
To demonstrate the behavior of the function $ \Sigma $  obtained by different approaches and compare them all together, we build two types of graphs. The first one
shown in Fig. \ref {allloop6} contains four different curves
corresponding  to four different approaches.
The second graph is a three-dimensional plot shown in Fig.\ref{3d6}. Here we plot the PT approximation, the ladder approximation and the second ladder approximation.
The first graph shows that all curves have almost the same behaviour. Analytically, it is perfectly described by a ladder approximation. This is also confirmed by the three-dimensional graph shown in Fig.\ref{3d6}. The inclusion of the second ladder does not change the solution qualitatively but provides a better match with PT. The function $ \Sigma $ has no restrictions for $ x \to \infty $ ($ \epsilon \to 0 $) for $ s> 0 $ and tends to a fixed point when $ s <0 $. This limit would correspond to the removal of the UV regularization. It can be seen that the summation of the entire infinite series does not lead to a finite theory. \\

{\bf D=8}

In the case of D=8, the PT series starts already from one loop and has the form

\beqa
\Sigma_{PT}(s,t,z)&=& \frac{z}{6}+\frac{s^2 + t^2}{144}z^2 +\frac{15 s^4 - s^3 t + s^2 t^2 - s t^3 + 15 t^4}{38880} z^3+\label{pt8}    \\
&+& \frac{8385 s^6 - 268 s^5 t + 206 s^4 t^2 - 192 s^3 t^3 + 206 s^2 t^4 -
   268 s t^5 +8385 t^6}{391910400}z^4  + ...      \nonumber
\eeqa
For $t=s$ the [7/6] Pade  approximant is
\beqa
\Sigma_{Pade}(x)=&& \frac{1}{s^2}\frac{0.17 x - 0.017 x^2 + 0.00040 x^3 + 0.000014 x^4 - 7.1\cdot10^{-7} x^5 +}{1 - 0.19 x + 0.014 x^2 -
 0.00046 x^3 + 6.9\cdot10^{-6} x^4 - 1.5\cdot10^{-8} x^5 -} \rightarrow  \nonumber \\
&& \leftarrow \frac{ +
 7.5\cdot10^{-9} x^6 + 1.2\cdot10^{-10} x^7}{ - 5.5\cdot10^{-10} x^6},
 \label{pade8}
\eeqa
where now  $x=zs^2$.

\begin{figure}[htb!]
\begin{center}
\leavevmode
\includegraphics[width=0.5\textwidth]{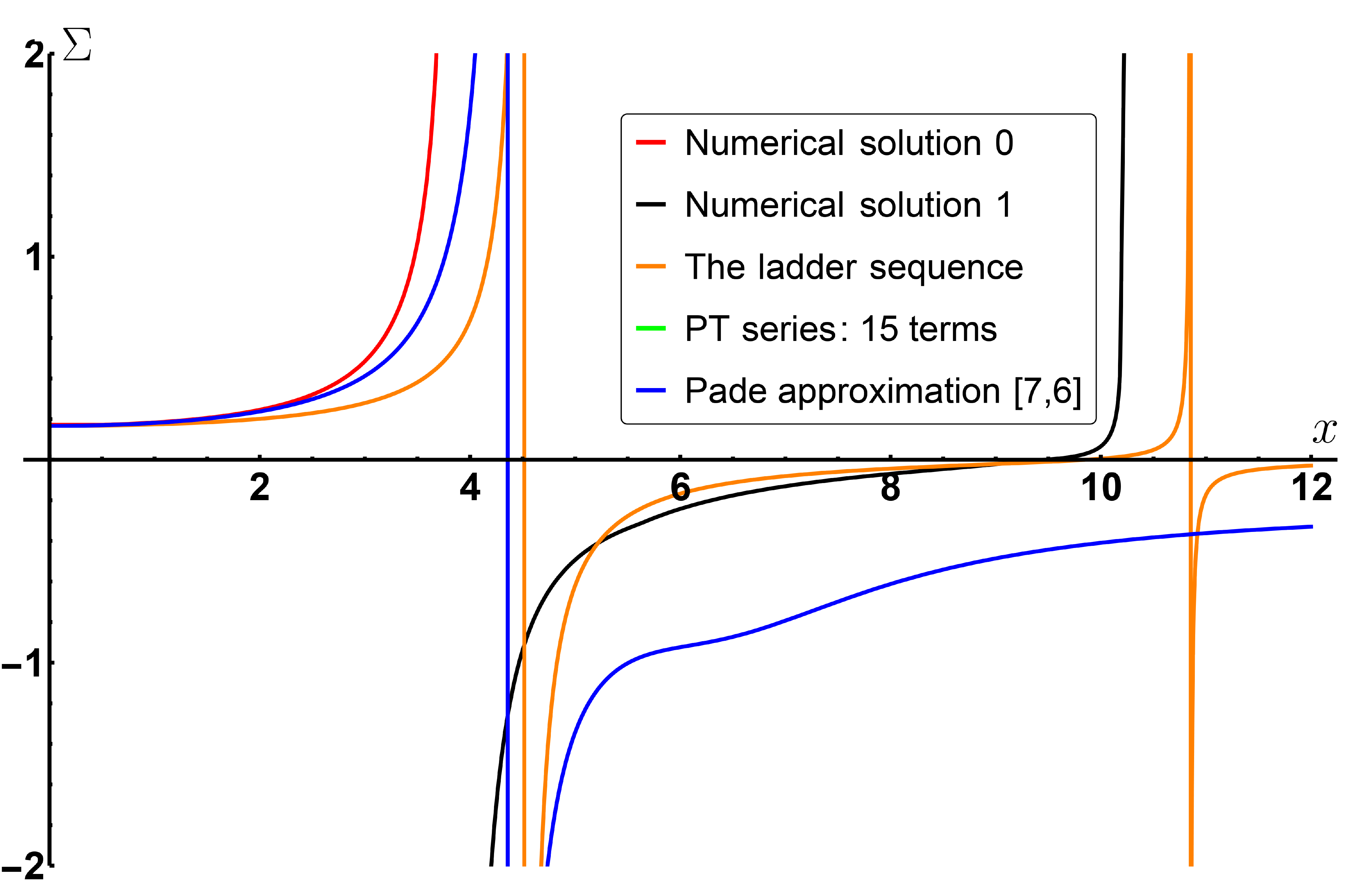}
\end{center}
\caption{Comparison of various approaches to solve eq.(\ref{eq8}) . The red and black lines are the numerical solutions described in the previous section between the first pole and between the first and the second ones. The green one is the PT. The blue one is the Pade approximation. And the last one is yellow which represents the Ladder analytical solution}
\label{allloop8}
\end{figure}

The ladder approximation is given by eq.(\ref{lad8}). The numerical solution starts with $ z = 0 $ and continues to the first pole $ z = z_1 $. Then we start it again for $ z> z_1 $ and reach the second pole at $ z = z_2 $ and so on. The poles coincide with the poles in the ladder approximation (\ref {lad8}).
A comparison of various curves for $t=s$ is shown in Fig.\ref{allloop8}.

It can be seen that in the first interval all  curves almost coincide. The PT curve exists only in the interval below the first pole. The Pade curve reproduces the first pole but does not fit to the others. The numerical curve reproduces both poles and is close to the ladder approximation.

\begin{figure}[htb!]
\begin{center}
\leavevmode
\includegraphics[width=0.3\textwidth]{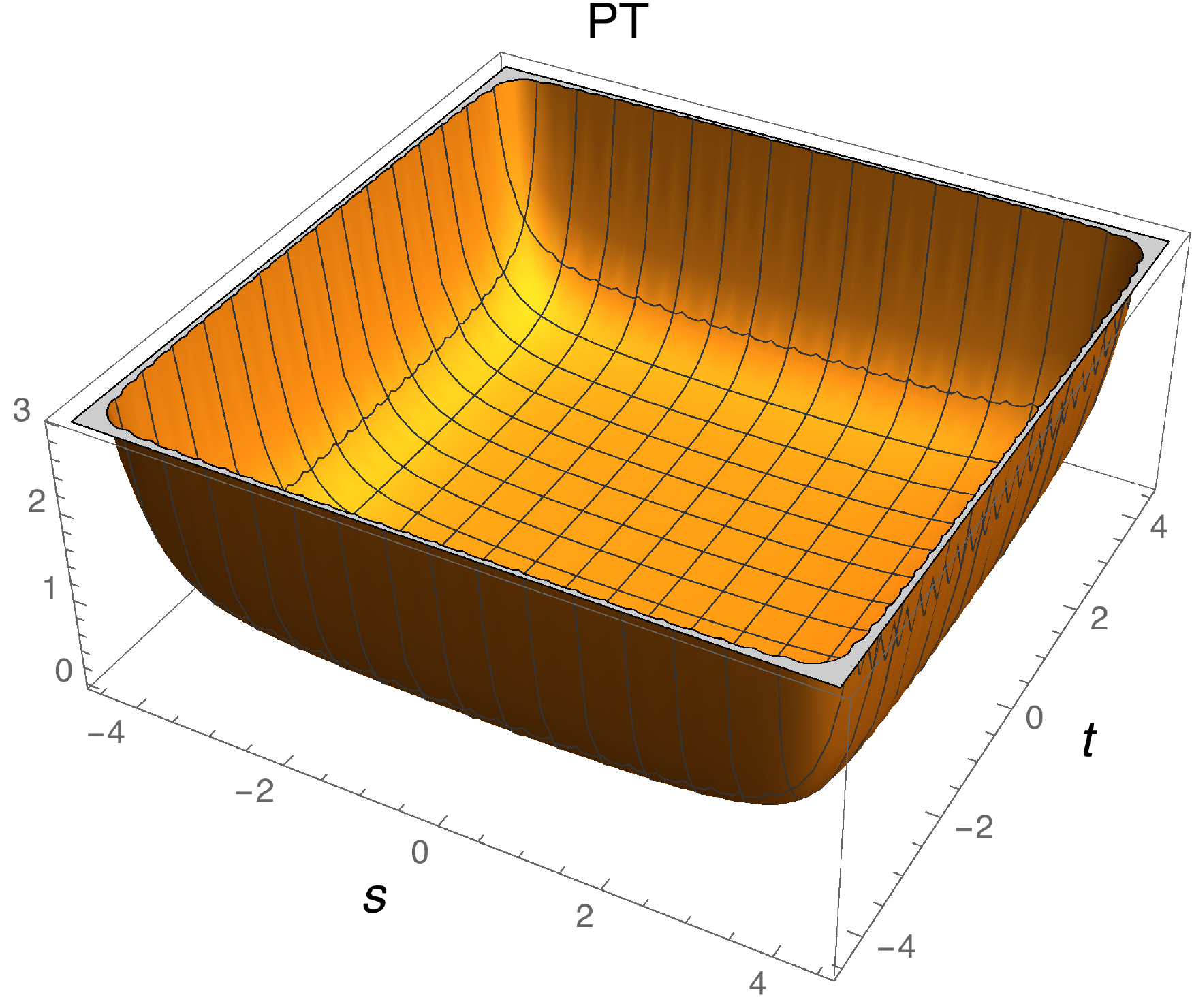}
\includegraphics[width=0.3\textwidth]{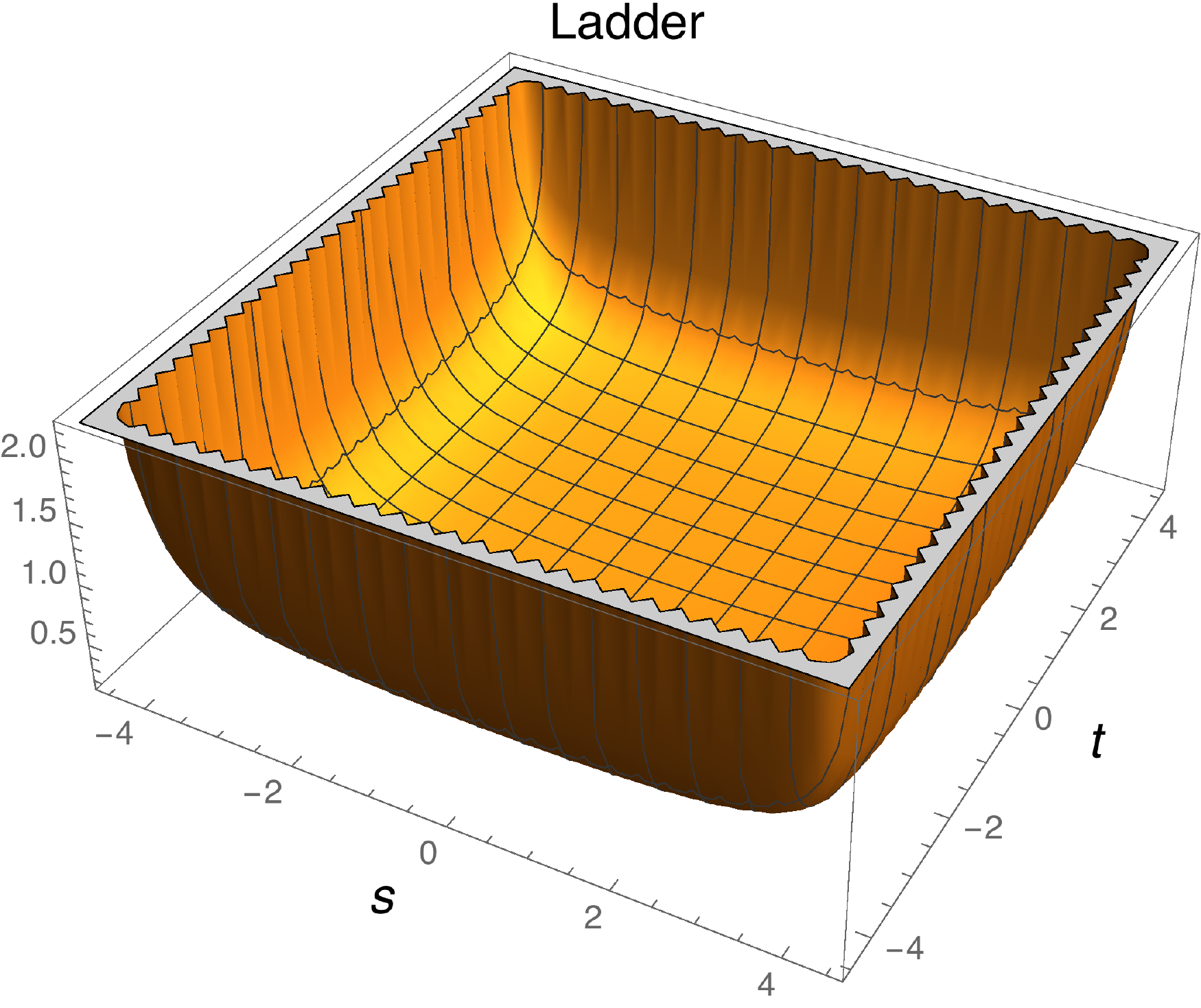}
\includegraphics[width=0.3\textwidth]{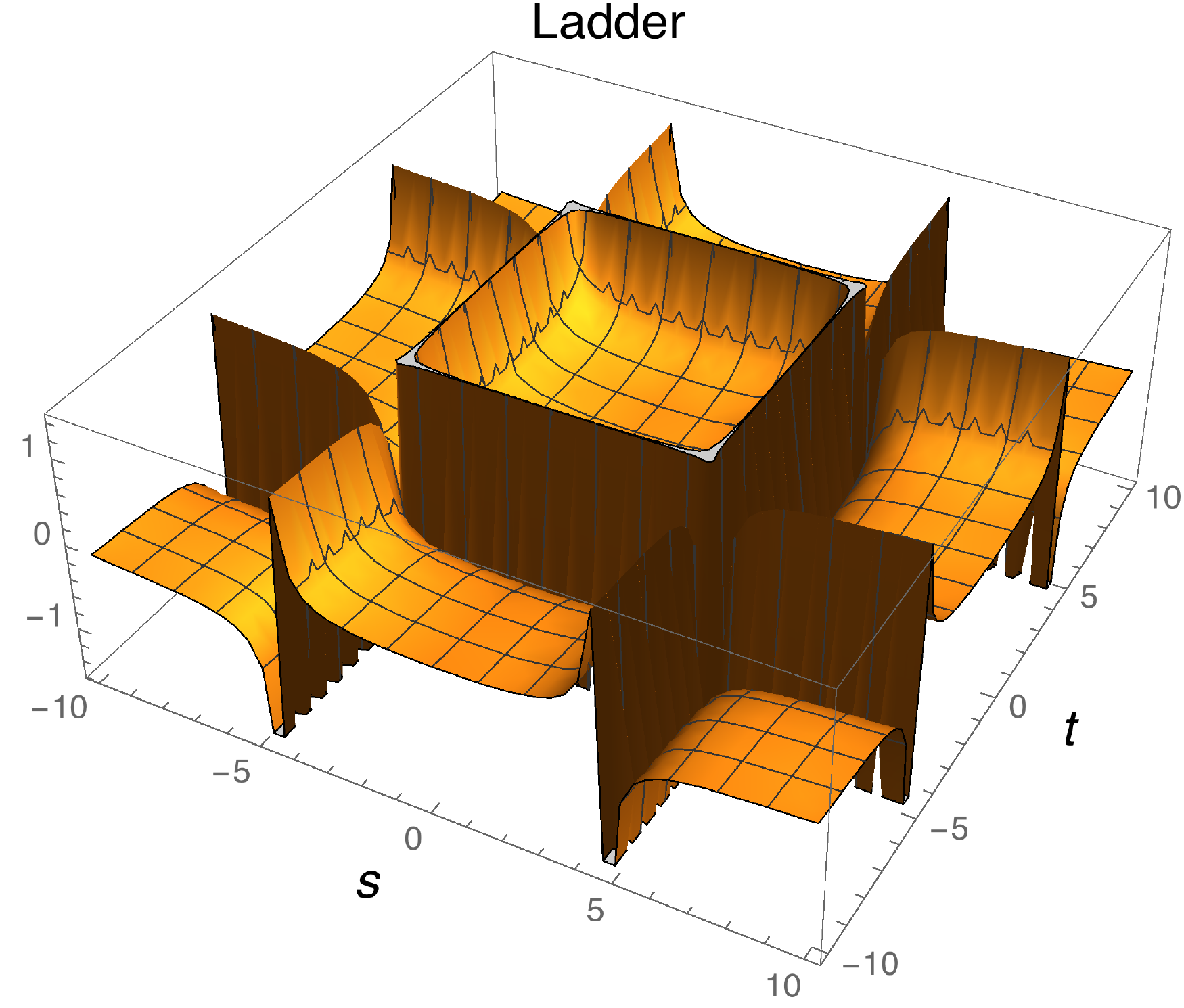}
\end{center}
\hspace{3cm} a \hspace{4cm} b \hspace{5cm} c
\caption{Comparison of PT(a)  and the  ladder approximation (b) in the region up to the first pole. The last plot (c) shows the ladder approximation beyond the first pole. One can clearly see the pole structure of the function $\Sigma$. }
\label{3d8}
\end{figure}

We present also the 3-dimensional plot in the $s-t$ plane for $z=1$ in Fig.\ref{3d8}.

It can be seen that the ladder as in the case of $ D = 8 $ gives a very accurate approximation to PT and allows you to go beyond the limits of the first pole. A comparison of the ladder approximation and the numerical solution of the complete equation confirms our conclusion that the ladder approximation gives the correct behavior of the function.

Again, we must admit that the limit $ z \to \infty $ ($ \epsilon \to 0 $) does not exist. The function has an infinite number of periodic poles for any choice of kinematics. Therefore, finiteness is not achieved when the sum over the entire cycles is taken into account. \\

{\bf D=10}

This case is quite similar to the $D=8$ one.  The PT series is now
\beqa
&&\Sigma_{PT}(s,t,z)  = \frac{(s + t)z}{120} + \frac{(4 s^4 + s^3 t + s t^3 + 4 t^4)z^2}{302400} + \\ \label{pt10}
&&+\frac{(2095 s^7 + 115 s^6 t + 33 s^5 t^2 - 11 s^4 t^3 - 11 s^3 t^4 +
  33 s^2 t^5 + 115 s t^6 + 2095 t^7)z^3}{68584320000} +  ...  \nonumber
\eeqa
while the [6/7] Pade approximation for $t=s$ reads
\beqa
\Sigma_{Pade}(x)=&& \frac{1}{s^2}\frac{0.017 x + 0.00025 x^2 + 6.5\cdot10^{-7} x^3 - 5.7\cdot10^{-10} x^4 -
 }{1 + 0.013 x +
 9.4\cdot10^{-6} x^2 - 1.1\cdot10^{-7} x^3 - 7.2\cdot10^{-11} x^4 +} \rightarrow  \nonumber \\
&& \leftarrow \frac{  - 2.1\cdot10^{-12} x^5 +  2.6\cdot10^{-16} x^6 + 7.3\cdot10^{-19} x^7}{+ 1.9\cdot10^{-13} x^5 - 6.4\cdot10^{-17} x^6 + 4.6\cdot10^{-21} x^7},
 \label{pade10}
\eeqa
where $x=zs^3$.

\begin{figure}[htb!]
\begin{center}
\leavevmode
\includegraphics[width=0.5\textwidth]{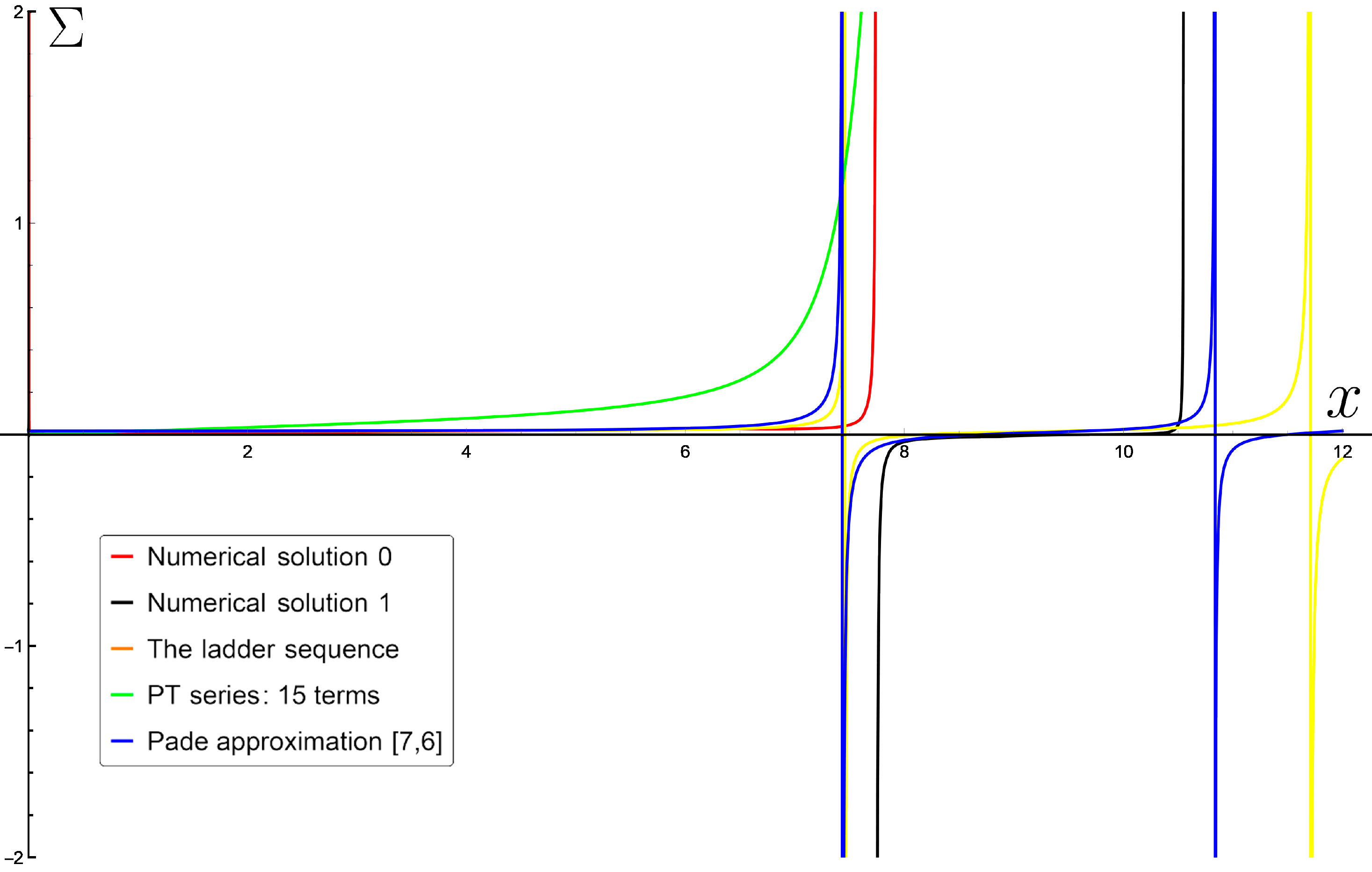}
\end{center}
\caption{Comparison of various approaches to solve eq.(\ref{eq10}) . The red and black lines are the numerical solutions described in the previous section before the first pole and between the first and the second ones. The green one is the PT. The blue one is the Pade approximation. The yellow one represents the Ladder analytical solution}
\label{allloop10}
\end{figure}

The ladder approximation is given by equations (\ref {lad10}, \ref {ladd}), and the numerical approximation  is constructed first for the interval from $ z = 0 $ to the first pole, and then continues to the second, etc.
The comparison of all the curves is shown in Fig.\ref{allloop10}.

The situation here is the same as in the case of $ D = 8 $. The ladder approximation works quite well, and its analytical solution qualitatively describes all the features of the full equation. The function $ \Sigma $ obeys an infinite number of periodic poles and one single pole is obtained from $ \Delta $ (\ref {ladd}). There is no limit when $ \epsilon \to 0 $.

Since the removal of the regularization ($ \epsilon \to 0 $) does not lead to finite amplitudes, it is necessary to perform a kind of renormalization procedure.  It has some peculiarities because these theories are nonrenormalizable.

\section{The renormalization procedure}

All higher dimensional gauge theories are non-renormalizable.  Of course, the scattering amplitudes can be made finite by subtracting  all UV divergences in some way. This is not a problem. The problem is that  the counter terms do not repeat the original Lagrangian and one gets new structures with increasing power of momenta at each step of perturbation theory. 
This means that subtracting the UV divergence each time, one has to define the normalization of a new operator, thus having a new arbitrary constant.  The number of these constants is infinite.  However, as we have demonstrated earlier, all the higher order divergences are related via the generalized RG equations. Hence,  changing the subtraction condition at a given loop, one consistently changes the normalization condition of an infinite set of operators.  Hence, one may hope to relate them removing the arbitrariness. 

\subsection{The scheme dependence}

All our consideration so far was based on the minimal subtraction scheme. To trace what happens when one changes the normalization condition, we consider now the non-minimal subtraction scheme. As an example we take the $D=8$ case where divergences appear already in one loop.

 Obviously, the leading divergences are scheme independent but the subleading ones depend on a scheme.  However, this dependence in all orders of PT is defined by a single arbitrary constant which appears in subtraction of a single one-loop box-type diagram. The recurrence relations obtained above are scheme independent. Indeed,  if one chooses the one-loop counter term in the form 
\beq
A_1'+B_{s1}'=\frac{1}{6\epsilon}(1+c_1 \epsilon)
 \eeq
 ($c_1=0$ corresponds to the minimal subtraction scheme), then using the recurrence relations for the subleading divergences, one gets the following additional term to the sum of the counter terms in all orders of PT (remind the notation $z\equiv g^2s^2/\epsilon$)
  \beq	
 \Delta \Sigma_{sB}'=c_1 z\frac{d\Sigma'_A}{dz}.
 \label{add}
 \eeq
Thus, the arbitrariness in the counter terms with an infinite number of derivatives is reduced in the leading order to the choice of the single parameter $c_1$. 
It is equivalent to a finite change of the renormalization constant $Z_4$
 \beq	
 Z_4=1+g^2s^2 c_1.
 \label{add1}
 \eeq
This is exactly what happens to renormalizable interactions except that there we redefine a single coupling  $g^2$ and here it is  an infinite series of couplings with increasing power of derivatives. 

Consider now what happens in the  sub-subleading order. 
In this case, the dependence on the subtraction scheme is contained also in the two-loop box-type diagram. Following the subleading case, we choose the counter term in the form 
\begin{equation}
A_2'+B_2'=\frac{s}{3!4!\epsilon^2}\left(1-\frac{5}{12}\epsilon+2c_1\epsilon+c_2\epsilon^2\right),
\end{equation}
where $c_1$ comes from the one-loop counter term and $c_2$ is the new subtraction constant. Using the recurrence relation for the sub-subleading divergences, one gets the following additional term  proportional to $c_2$ in all orders of PT:
\beq	
 \Delta \Sigma_{sC}'=c_2 z^2\frac{d\Sigma_A'}{dz}.
 \eeq
This corresponds to the finite renormalizations
\beq
Z_4=1+g^2s^2 c_1+g^4s^4c_2
 \label{add2}
 \eeq
This simple pattern obviously has a one-loop origin since it comes from the leading divergences and they are defined by the one-loop box diagram.

The situation with dependence on $c_1$ in the sub-subleading order  is more complicated. There are two contributions here: the linear and quadratic. The quadratic dependence obviously appears from the substitution of expression (\ref{add1}) into the
minimal scheme counter term $\Sigma'_A$ (\ref{add}), which gives the second derivative of $\Sigma'_A$.  However, the redefinition of the coupling contains  an extra part  compared to (\ref{add2}) which is proportional to  $c_1^2$ that gives the first derivative of $\Sigma'_A$. All together the full quadratic dependence has the form
\beq 
 \Delta \Sigma_{sC}'=-c_1^2\frac{z}{4!}\left(\frac{d\Sigma'_A}{dz}-12 \frac{d^{2}\Sigma'_A}{dz^{2}}\right).\label{two}
\eeq
Using the recurrence relations in the sub-subleading order, we have checked that this result is valid in all orders of PT.

The situation with the linear term is not that straightforward. It is not given by the leading term only but involves also the subleading one.  Here for the first time we meet the situation where the renormalization is not reduced to a simple multiplication.  It happens because the subleading terms depend on both $s$ and $t$ and one cannot separate them anymore.

\begin{figure}[htb]
\begin{center}
\leavevmode
\includegraphics[width=0.8\textwidth]{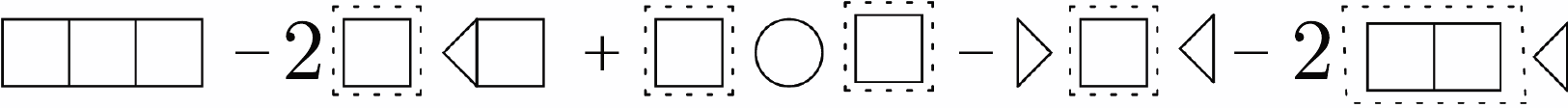}
 \caption{$\R'$-operation for the 3-loop box diagram}\label{3_loop}
 \end{center}
 \end{figure}

This is clearly seen in the third order of PT. Namely, if we consider the $\R'$-operation of the 3-loop box diagram and calculate the {\em arbitrariness} $\Delta \Sigma_{sC}'$ which is due to arbitrariness in the two loop counterterm (the last diagram in Fig.\ref{3_loop})
the latter is independent of the $t$ contribution. The reason is that while the two loop box contains the $t$ contribution in the subleading order, the arbitrariness is contained only in the $s$ term. At the same time, when one evaluates the sub-subleading {\em divergence}  in the 3-loop box diagram using the $\R'$-operation, one has a nonzero contribution from both the $s$ and $t$ terms in the last diagram in Fig.\ref{3_loop}. The two expressions are obviously unrelated
 
\beq	
\Delta \Sigma_{sC}'(3-loop) = - \frac{719 c_1 s^2}{1036800 \epsilon}, 
\label{delta_3}
\eeq
 whereas $\Sigma_{sB}'$ in 3 loops has the following form:
\beq 
\Sigma_{sB}'(3-loop) = - \frac{71 s^2}{345600\epsilon^2}.\label{sB_3}
\eeq

The discrepancy comes  from this last term in Fig.\ref{3_loop}. To see this, we subtract the unmatched $t$ contribution from $\Sigma_{sB}'$ and compare it with $\Delta \Sigma_{sC}'$. We call it $\Sigma_{sB}^{'trunc}$
\beq 
 \Sigma_{sB}^{'trunc}(3-loop) = - \frac{719 s^2}{3110400 \epsilon^2}. \label{t_prime}
\eeq
Taking the derivative with respect to $z$, one reproduces the desired result
\beq	
 \Delta \Sigma_{sC}'(3-loop)=c_1 z\frac{d\Sigma_{sB}^{'trunc}}{dz}(3-loop).
 \eeq
The situation repeats itself in the fourth order of PT being even more tricky. 

This consideration shows us what actually goes wrong in the renormalization procedure and suggests the right way to formulate it. The key reason is that the divergent expressions and, hence, the counter terms
are not constants anymore but are polynomials of momenta. This means that this momentum dependent counter terms have to enter {\it inside} the integrals when performing the $\R$-operation, i.e. this is not a simple multiplication procedure anymore.

\subsection{Kinematically dependent renormalization}
 
Based on the performed analysis, below we describe how the renormalization procedure can be reformulated in non-renormalizable theories and illustrate it  by the example of two and three loop divergences in D=8 SYM theory. Formally, it looks {\it precisely} like a familiar renormalization procedure in any renormalizable theory, but the renormalization constant $Z$ becomes the function of kinematic variables  and acts on the amplitude not as a simple multiplication but as the operator in momentum space.  Namely, to remove all the UV divergences in the amplitudes and get a finite answer, one follows the usual prescription multiplying
the bare amplitude by the proper renormalization constant and replacing the bare coupling 
with the renormalized one:
\beqa
\bar{\A}_4& = &Z_4(g^2) \bar{\A}_4^{bare}\big|_{g^2_{bare}\mapsto g^2Z_4}, \label{mult}\\
g^2_{bare}&=&\mu^\epsilon Z_4(g^2)g^2, \label{coupling}
\eeqa
where $\bar{\A}_4$ is the ratio ($\mathcal{A}_4/\mathcal{A}_4^{(0)}$).
Remind also that the renormalization constant $Z_4$ can be calculated diagrammatically with the help
of the following standard operation~\cite{Rop}:
\beq
Z  = 1-\sum_i\K\R' G_i. \label{ZZ}
\eeq

The essential difference between the non-renormalizable and the renormalizable cases manifests itself in momentum dependence of the renormalization constant $Z$. This actually means that it becomes the operator acting on the amplitude according to the rules of the $\R$-operation. 
To demonstrate  how this  renormalization procedure works, we apply eqs. (\ref{mult},\ref{coupling}) to the singular part of the amplitude 
\beq
\bar{\A}_4= 1-\frac{g^2 st}{3!\epsilon}-\frac{g^4 st}{3!4!}\left(\frac{s^2+t^2}{\epsilon^2}+\frac{27/4 s^2+1/3 st +27/4 t^2}{\epsilon}\right) + ...  \label{gamma}
\eeq
order by order in PT.

In the one loop order the coupling is not changed   $g^2_{bare}=\mu^\epsilon g^2$
and the renormalization constant is chosen in the form $Z_4=1+\frac{g^2 st}{3!\epsilon}$. This leads to a finite answer. Notice that the renormalization constant is not really a constant but depends on the kinematic factors $s$ and $t$! 

In the two loop order the coupling is changed now according to (\ref{coupling}), namely,
\beq
g^2_{bare}=\mu^\epsilon g^2(1+\frac{g^2 st}{3!\epsilon}) \label{1lcoup}\eeq
\\
 and the renormalization constant is taken in the form
 \beq
Z_4=1+\frac{g^2 st}{3!\epsilon}+\frac{g^4 st}{3!4!}\left(\frac{A_2s^2+B_2st+A_2t^2}{\epsilon^2}+
\frac{A_1s^2+B_1st+A_1t^2}{\epsilon}\right),\label{2lz}
\eeq
where the coefficients $A_i$ and $B_i$  have to be chosen in a way to cancel all divergences both local and nonlocal ones.  

\begin{figure}[!ht]
\begin{center}
\includegraphics[scale=0.5]{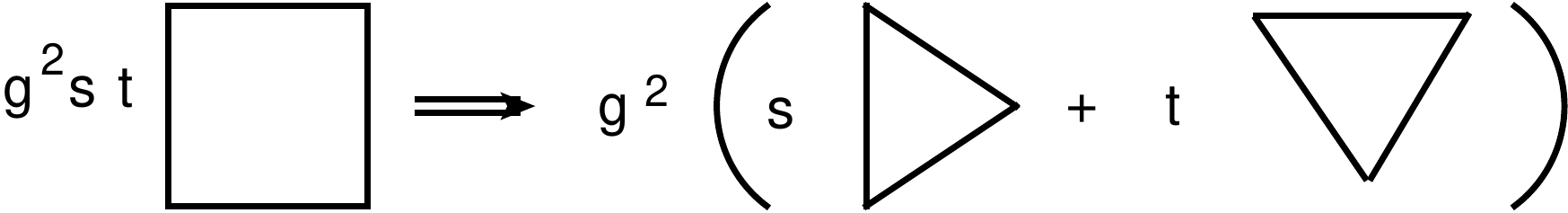}
\caption{Action of the Z-operator at the two loop level}
\label{action}
\end{center}
\end{figure}

Consider how it works in practice:  When substituting eqs.(\ref{1lcoup},\ref{2lz}) into eq.(\ref{mult}), one can notice that the replacement of $g^2_{bare}$ by expression (\ref{1lcoup}) in the one loop term ($\sim g^2$) and multiplication of one loop contributions from the renormalization constant $Z_4$  and from the amplitude $\bar{\A}_4$ have the effect of subtraction of subdivergences in the two loop graph. This is exactly what guarantees the locality of the counter terms within the $\R$-operation. However, contrary to the renormalizable case, here the renormalization constant contains the kinematic factors, the powers of momenta, which are external momenta for the subgraph but are internal ones for the whole diagram. This means that evaluating the counterterm they have to be inserted inside the remaining diagram and integrated out. To clarify this point, we consider the corresponding term which appears when multiplying the one loop Z factor by the one loop amplitude. The $s$ and $t$ factors from the Z factor have to be inserted inside the box diagram,
as shown in Fig.\ref{action}

This means that the usual {\it multiplication} procedure has to be modified: the Z factor becomes the {\it operator} acting on the diagram which inserts the powers of momenta inside the diagram. This looks a bit artificial but exactly reproduces the $\R$-operation for the two loop diagram shown below
in Fig.\ref{2loop}.
\begin{figure}[!ht]
\begin{center}
\includegraphics[scale=0.45]{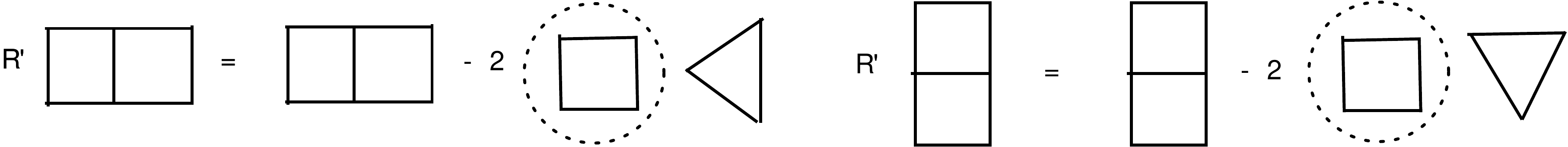}
\caption{$\R'$-operation for the two loop diagrams}\label{2loop}
\end{center}
\end{figure}

Thus, inserting eqs.(\ref{2lz},\ref{1lcoup}) into eq.(\ref{mult}) and having in mind that 
\beq
sTriangle=-\frac{s}{4!\epsilon}(1+\frac{19}{6}\epsilon), \ \  tTriangle=-\frac{t}{4!\epsilon}(1+\frac{19}{6}\epsilon),
\eeq
one gets
\beqa
\bar{\A}_4& = &Z_4(g^2) \bar{\A}_4^{bare}|_{g^2_{bare}\mapsto g^2Z}\nonumber \\
&=& 1-\frac{g^2 \mu^\epsilon st}{3!\epsilon}+\frac{g^2 st}{3!\epsilon}-\frac{g^4 \mu^{2\epsilon} st}{3!4!}\left(\frac{s^2+t^2}{\epsilon^2}+\frac{27/4 s^2+1/3 st +27/4 t^2}{\epsilon}\right)  \\
&+& 2\frac{g^4 st}{3!\epsilon} \mu^{\epsilon}\frac{s^2+t^2}{4!\epsilon}(1+\frac{19}{6}\epsilon)+
\frac{g^4 st}{3!4!}\left(\frac{A_2s^2+B_2st+A_2t^2}{\epsilon^2}+
\frac{A_1s^2+B_1st+A_1t^2}{\epsilon}\right).\nonumber
\eeqa

One can see that the one loop divergences ($\sim g^2$) cancel and the cancellation of the two loop ones requires
\beqa
\frac{1}{\epsilon^2}: && -\frac{s^2+t^2}{3!4!}st +2 \frac{s^2+t^2}{3!4!}st + \frac{A_2s^2+B_2 st +A_2 t^2}{3!4!}st=0,\nonumber \\
\frac{\log\mu}{\epsilon}: &&  -2\frac{s^2+t^2}{3!4!}st +2 \frac{s^2+t^2}{3!4!}st =0,\nonumber \\
\frac{1}{\epsilon}: &&-\frac{st}{3!4!}(\frac{27}{4} s^2+\frac 13 st +\frac{27}{4} t^2)+2\frac{st}{3!4!}(s^2+t^2)\frac{19}{6}
+\frac{st}{3!4!}(A_1s^2+B_1st+A_1t^2)=0. \nonumber
\eeqa
One deduces that $A_2=-1,B_2=0, A_1=\frac{5}{12}, B_1=\frac 13$, so that the renormalization constant $Z_4$ takes the form
\beq
Z_4=1+\frac{g^2 st}{3!\epsilon}+\frac{g^4 st}{3!4!}\left(-\frac{s^2+t^2}{\epsilon^2}+
\frac{5/12 s^2+1/3st+5/12 t^2}{\epsilon}\right),\label{2lzp}
\eeq
which exactly corresponds to the one obtained using eq.(\ref{ZZ}). This expression now has to be substituted into eq.(\ref{coupling}) to obtain the renormalized  coupling. Note that it also depends on kinematics. 

\begin{figure}[ht]
\begin{center}
\includegraphics[scale=0.5]{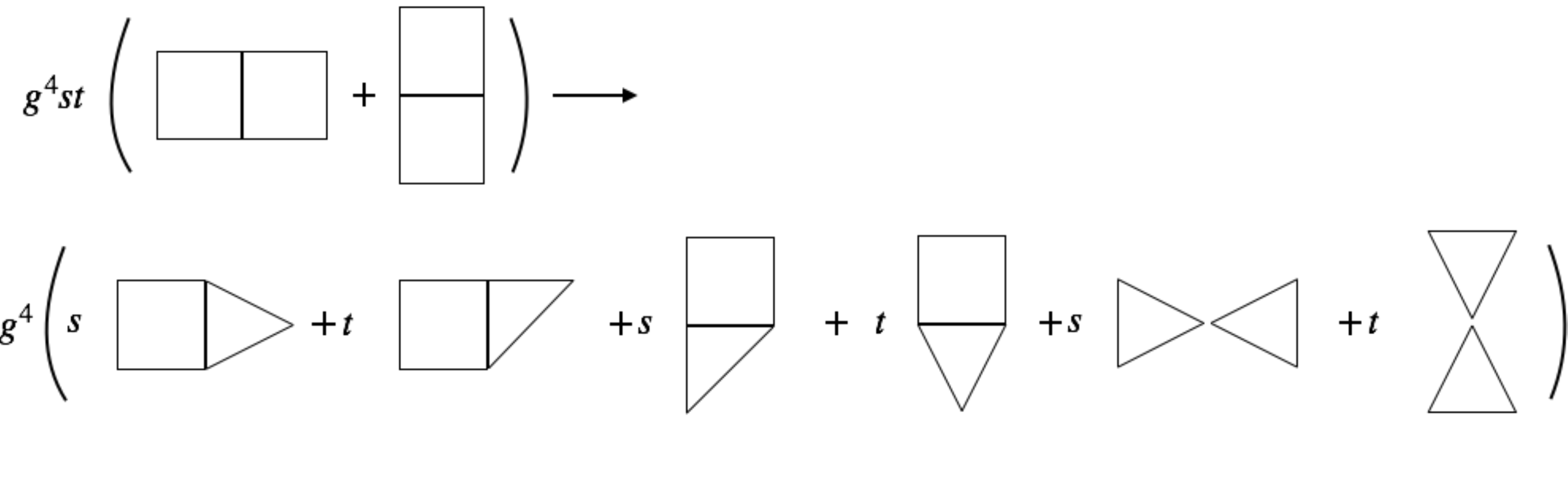}
\caption{Action of the Z-operator at the three loop level. The first, second and the last two diagrams in the r.h.s correspond to the three loop box counter terms and the third and fourth ones to the tennis-court counter terms.}\label{3loop}
\end{center}
\end{figure}

The same way one can trace the action of the Z-operator in the three-loop diagram, as is shown in Fig.\ref{3loop}. In this case, besides the  3-loop box diagram one also has the tennis-court one, and the resulting counterterms correspond to both of them.

In the context of the present discussion the transition to a non-minimal scheme is equivalent to the multiplication of the amplitude by the finite renormalization constant
 \beq z = 1+g^2 s t c_1 \label{fin}
 \eeq
 and the corresponding finite change of the coupling $g$.
 This looks similar to the renormalizable case though the meaning is different. Again, it is not simply the multiplication  but  the action of the operator which is also kinematically dependent.  Therefore, it is not a simple change of a single coupling but  of the whole infinite series of higher derivative terms.
 
 Similarly, the subtraction arbitrariness of the double box influences the subsubleading divergences and results in higher order terms in eq.(\ref{fin}) like in eq.(\ref{2lzp}),  just as in renormalizable theories~\cite{we4}.  
 Therefore, the whole arbitrariness is accumulated in one renormalization constant evaluated order by order in PT, which acts as an operator and generates an infinite series of terms.
 
In fact, this means that we build this way an induced higher derivative theory where higher terms appear order by order of PT with fixed coefficients. For instance, the one loop term
$g^2 s t/\epsilon$ generates the gauge invariant counter term 
$$ \frac{g^2}{\epsilon}D_\rho D_\lambda F_{\mu\nu}D_\rho D_\lambda F_{\mu\nu},$$
that contains higher derivatives as well as new vertices with extra gauge fields, etc.

\section{High energy behaviour}

Assuming that one accepts these arguments, there is still a problem that at each order of PT the amplitude increases with energy, thus violating the unitarity.  However, apparently, this problem has to be addressed after summation  of the whole PT series. While each term of PT behaves badly, the whole sum might behave differently. 

To analyze the high energy behaviour of the full amplitude, one can use the solutions of the generalized RG equations obtained above. Indeed, like in any renormalizable case the high energy behaviour is associated with the UV divergences. Considering the case when $s\sim t\sim u\sim E^2$ and expanding the amplitude over $\epsilon$, one finds the one-to-one correspondence between the coefficient of the leading pole $(g^2 E^{(D-4)})^n/\epsilon^n$ and the leading asymptotic term $(g^2E^{(D-4)})^n \log^n E^2$: $g^2/\epsilon \leftrightarrow - g^2 \log E^2$ Thus, $g^2/\epsilon=z\to -\infty$, which we considered in Sec.4, corresponds to the limit $E\to\infty$.

Having this in mind we can analyze the high energy asymptotics of the amplitudes in $D=6,8$ and $10$.
The task becomes more complicated since we have the function of two variables and one may have different limits in different directions. We looked for the case when $s>0$ and $t,u<0$ corresponding to the c.m. frame. 

 In the D=6  case, in the leading order the full amplitude behaves qualitatively like the ladder.
 For the $s-t$ partial amplitude it contains the exponent $exp[-(s+t)]\log{E}$, and since $s+t=-u>0$, one has a decreasing exponent and hence a smooth function of energy without violation of unitarity.
 The same is true for the $s-u$ partial amplitude. However, for the $u-t$ amplitude one has $u+t=-s<0$, which results in increasing exponent. This amplitude obviously violates unitarity and  spoils the picture. The subleading asymptotics does not improve the situation having the same type of behaviour.

In the D=8 and 10 cases, again the ladder diagrams qualitatively correctly reproduce the behaviour of the full amplitude. Here all partial amplitudes behave similarly. They have  poles for finite values of $z$ and, hence, for finite values of $E$.  This is similar to QED; however, in this case the pole is much closer due to the power law behaviour of the function. And there are multiple poles. Thus, again one has problems with unitarity at high energy.

It would be interesting to find an example of a theory where such kind of summation leads to a  smooth high energy limit like in asymptotically free theory in the renormalizable case.

\section{Conclusion}

Our main concern here was the understanding of the structure of UV divergences in supersymmetric gauge theories with maximal supersymmetry. 

We restricted ourselves to the on-shell scattering amplitudes since after all it is the S-matrix, which we want to make finite. 

Our main results can be formulated as follows: 

1) The on-shell scattering amplitudes contain the UV divergences that start from one loop (three loops) and do not cancel (except for the all loop cancellation of the bubbles and triangles). 

2) These divergences possess increasing powers of momenta (derivatives) with increasing order of PT.  For the four-point scattering amplitude this manifests itself as increasing power of the Mandelstam variables $s$ or $t$.  This means that the theory is not renormalizable by power counting. 

3) Nevertheless, all the higher loop divergences are related to the lower ones via explicit pole equations which are the generalization of the RG equations to the case of non-renormalizable theories. The leading divergences are governed by the one-loop counter term, the subleading ones - by the two-loop counter term, etc. This happens exactly as in the well known case of renormalizable interactions.

4) The summation of the leading and subleading divergences can be performed by solving the generalized RG equations. The solution to these equations depends on dimension and has a different form in different dimensions. For particular sets of diagrams one can get an analytical solution, while in the general case it is only numerical. 

5) In D=6 the solution is characterized by the exponential function which decreases for some partial amplitudes and increases for the other
as a function of  $z=g^2/\epsilon$.
In D=8 and D=10 the solutions  possess an infinite number of poles.  
This means that they do not have a finite limit when $z\to\infty$ ($\epsilon\to 0$) which would correspond to the finite answer when removing the regularization.

6) We reformulate the multiplicative renormalization procedure with replacement of the renormalization constant by an operator that depends on kinematics. 
As a result, one can construct a higher derivative theory that gives finite scattering amplitudes with a single arbitrary coupling $g$ defined in PT within a given renormalization scheme. Transition to another scheme is performed by  the action on the amplitude of the finite renormalization operator. 

7) The high energy behaviour of the amplitudes is governed by the generalized RG equations just as in renormalizable theories. In the three examples, which we considered, this behaviour is different but in all the cases the amplitudes either increase with energy or hit the pole at finite energy like in QED.

8) Thus, the maximal supersymmetric gauge theories at higher dimensions despite many attractive features still happened to be inconsistent at high energies. We hope that the methods of analysis developed here can be used in  other non-renormalizable theories including gravity.

\section*{Acknowledgements}
This work was supported by the Russian Science Foundation grant \# 16-12-10306. Arthur Borlakov is grateful to the Russian Foundation for Basic Research for supporting the grant \# 17-02-00872.


\begin{thebibliography}{999}

\bibitem{Reviews_methods}
Z.~Bern, Yu-tin ~Huang, Basics of Generalized Unitarity. {\em J.Phys. A} {\bf 2011}, {\em 44}, 454003, doi:10.1088/1751-8113/44/45/454003.\\
H.~Elvang, Yu-tin Huang, Scattering Amplitudes. {\bf 2013}, 268 pp, arXiv:1308.1697 [hep-th].

\bibitem{BDS4point3loop_et_all}
J. Bartels, V. Schomerus, M. Sprenger, The Bethe roots of Regge cuts in strongly coupled N=4 SYM theory. {\em JHEP} {\bf 2015}, {\em 1507}, 098, doi:10.1007/JHEP07(2015)098.\\
L.~J.~Dixon, J.~M.~Drummond, C.~Duhr, J.~Pennington, The
four-loop remainder function and multi-Regge behaviour at NNLLA in
planar $\mathcal{N}=4$ super-Yang-Mills theory. {\em JHEP} {\bf 2014}, {\em 1406}, 166, doi:10.1007/JHEP06(2014)116.

\bibitem{N=8SUGRA finiteness}
Z.~Bern, J.~J.~Carrasco, L.~Dixon, H.~Johansson, R.~Roiban, Amplitudes and Ultraviolet Behavior of $N=8$ Supergravity. {\em Fortsch.\ Phys.} {\bf 2011}, {\em 561}, doi:10.1002/prop.201100037.\\
R. ~Kallosh, 7(7) Symmetry and Finiteness of N = 8 Supergravity. {\em JHEP} {\bf 2012}, {\em 1203}, 083, doi:10.1007/JHEP03(2012)083.\\
Paul Heslop, Arthur E. Lipstein, On-shell diagrams for N = 8 supergravity amplitudes. {\em JHEP} {\bf 2016}, {\em 1606}, 069, doi:10.1007/JHEP06(2016)069 [hep-th].

\bibitem{GeneralDimensions}
R.~H.~Boels, D.~O'Connel, Simple superamplitudes in higher
dimensions. {\em JHEP} {\bf 2012}, {\em 1206}, 163,
doi:10.1007/JHEP06(2012)163.

\bibitem{Reviews_Ampl_General}
Andrei Smilga, Ultraviolet divergences in non-renormalizable supersymmetric theories. {\em Phys.Part.Nucl.Lett.} {\bf 2016}, {\em 14},  no.2, 245-260, doi:10.1134/S1547477117020315.\\
J. Broedel, M. Sprenger, Six-point remainder function in multi-Regge-kinematics: an efficient approach in momentum space. {\em JHEP} {\bf 2016}, {\em 1605}, 055, doi:10.1007/JHEP05(2016)055.\\
T.~Dennen, Yu-tin~Huang, Dual Conformal Properties of Six-Dimentional Maximal Super Yang-Mills Amplitudes. {\em JHEP} {\bf 2011}, {\em 1101}, 140, doi:10.1007/JHEP01(2011)140.

\bibitem{SpinorHelisity_extraDimentions}
S.~Caron-Huot, D.~O'Connel, Spinor Helicity and Dual Conformal Symmetry in Ten Dimensions. {\em JHEP} {\bf 2011}, {\em 1108}, 014, doi:10.1007/JHEP08(2011)014.\\
C. ~Cheung, D. ~O’Connel, Amplitudes and Spinor-Helicity in Six Dimensions. {\em JHEP} {\bf 2014}, {\em 0907}, 075, doi:10.1088/1126-6708/2009/07/075.

\bibitem{we0}
L.V. Bork, D.I. Kazakov, D.E. Vlasenko, On the amplitudes in N=(1,1) D=6 SYM. {\em JHEP} {\bf 2013}, {\em 1311}, 065, doi:10.1007/JHEP11(2013)065.

\bibitem{we1}
L.V. Bork, D.I. Kazakov, D.E. Vlasenko, Challenges of D=6 N=(1,1) SYM theory. {\em Phys. Lett.} {\bf 2014}, {\em B374}, doi:10.1016/j.physletb.2014.05.022.

\bibitem{we2}
L.V. Bork, D.I. Kazakov, M.V. Kompaniets, D.M. Tolkachev, D.E. Vlasenko, Divergences in maximal supersymmetric Yang-Mills theories in diverse dimensions. {\em JHEP} {\bf 2015}, {\em 1511}, 059, doi:10.1007/JHEP11(2015)059.

\bibitem{we3}
D.I. Kazakov, D.E. Vlasenko, Leading and subleading UV divergences in scattering amplitudes for D=8 SYM theory in all loops. {\em Phys.Rev.} {\bf 2017}, {\em D95}, no.4, 045006, doi:10.1103/PhysRevD.95.045006.

\bibitem{we4}
A.T. Borlakov, D.I. Kazakov, D.M. Tolkachev, D.E. Vlasenko, Summation of all-loop UV Divergences in Maximally Supersymmetric Gauge Theories. {\em JHEP} {\bf 2016}, {\em 1612}, 154, doi:10.1007/JHEP12(2016)154.\\
D.I. Kazakov, A.T. Borlakov, D.M. Tolkachev, D.E. Vlasenko, Structure of UV divergences in maximally supersymmetric gauge theories. {\em Phys.Rev.} {\bf 2018}, {\em D97}, no.12, 125008, doi:10.1103/PhysRevD.97.125008.

\bibitem{we5}
D.I. Kazakov, Kinematically dependent renormalization. {\em Phys. Lett.} {\bf 2018}, {\em B786}, 327-331, doi:10.1016/j.physletb.2018.10.002.

\bibitem{Sigel_D=6Formalism}
T.~Dennen, Yu-tin~Huang, W.~Siegel, Supertwistor space for D=6
maximal super Yang-Mills. {\em JHEP} {\bf 2010}, {\em 1104}, 127, doi:10.1007/JHEP04(2010)127.

\bibitem{PureSpinorsMarfa}
C.~R.~Mafra, O.~Schlotterer, 	
Two-loop five-point amplitudes of super Yang-Mills and supergravity in pure spinor superspace. {\em JHEP} {\bf 2015}, {\em 1510}, 124, doi:10.1007/JHEP10(2015)124.\\
C.~R.~Mafra, Pure Spinor Superspace Identities for Massless Four-point Kinematic Factors. {\em JHEP} {\bf 2008}, {\em 0804}, 093, doi:10.1088/1126-6708/2008/04/093.\\
C.~R.~Mafra, Superstring Scattering Amplitudes with the Pure Spinor Formalism. arXiv:0902.1552 [hep-th]. 

\bibitem{10dOnShell}
I.~Bandos, Spinor frame formalism for amplitudes and constrained superamplitudes of 10D SYM and 11D supergravity,
  {\em JHEP} {\bf 2018}, {\em 1811} 017, 
  doi:10.1007/JHEP11(2018)017
  [arXiv:1711.00914 [hep-th]].\\
I.~Bandos, An analytic superfield formalism for tree superamplitudes in D=10 and D=11,
  {\em JHEP} {\bf 2018}, {\em 1805} 103,
  doi:10.1007/JHEP05(2018)103,
  [arXiv:1705.09550 [hep-th]]. 

\bibitem{Bern:2005iz}
Z.~Bern, L.~J.~Dixon, V.~A.~Smirnov, 	
Iteration of planar amplitudes in maximally supersymmetric Yang-Mills theory at three loops and beyond. {\em Phys.\ Rev.\ D} {\bf 2005}, {\em 72}, 085001, doi:10.1103/PhysRevD.72.085001.\\
Z.~Bern, M.~Czakon, L.~J.~Dixon, D.~A.~Kosower, V.~A.~Smirnov, The Four-Loop Planar Amplitude and Cusp Anomalous Dimension in Maximally Supersymmetric Yang-Mills Theory. {\em Phys.Rev. D} {\bf 2007}, {\em 75}, 085010, doi:10.1103/PhysRevD.75.085010.\\
Z.~Bern, J.~J.~M.~Carrasco, H.~Johansson and R.~Roiban, The Five-Loop Four-Point Amplitude of N=4 super-Yang-Mills Theory. {\em Phys.Rev.Lett.} {\bf 2012}, {\em 109}, 241602, doi:10.1103/PhysRevLett.109.241602.

\bibitem{BPHZ}
 N. Bogoliubov and O. Parasiuk, {\em \"Uber die Multiplikation der Kausalfunktionen in der Quan- tentheorie der Felder}, Acta Math. 97 (1957) 227–266.\\
 K. Hepp, {\em Proof of the Bogolyubov-Parasiuk theorem on renormalization}, Commun. Math. Phys. 2 (1966) 301–326.\\
 W. Zimmermann, {\em Local field equation for A4-coupling in renormalized perturbation theory}, Commun. Math. Phys. 6 (1967) 161–188; W. Zimmermann, {\em Convergence of Bogoliubov’s Method of Renormalization in Momentum Space}, Comm. Math. Phys. 15 (1969) 208–234.\\
N. N. Bogolyubov, D.V. Shirkov, (1957, 1973, 1976, 1984)  \textit{Introduction to the Theory of Quantized Fields} [in Russian] (Moscow: Nauka)\\
English transl: (1980) \textit{Introduction to the Theory
of Quantized Fields, 3rd ed.} (New York: Wiley) \\	
O.I. Zavyalov, (1979) {\it Renormalized Feynman Diagrams} [in Russian], (Moscow: Nauka);
English transl.: (1990) {\it Renormalized Quantum Field Theory} (Dordrecht :Kluwer)

\bibitem{hooft}
G.'t ~Hooft, Dimensional regularization and the renormalization group. {\em Nuclear Physics B} {\bf 1973}, {\em 61}, 455-468, doi:10.1016/0550-3213(73)90376-3

\bibitem{Rop}
A. N. Vasiliev, Quantum Field Renormalization Group in Critical Behavior Theory and Stochastic Dynamics. (Petersburg Inst. Nucl. Phys., St. Petersburg State University, 1998); English transl: The Field Theoretic Renormalization Group in Critical Behavior Theory and Stochastic Dynamics (Chapman \& Hall/CRC, Boca Raton, 2004).
\end{thebibliography}
\end{document}